\documentclass[11pt]{article}%
\usepackage{graphics,mathrsfs}
\usepackage{epsfig}
\usepackage{graphicx}
\usepackage{amssymb}
\usepackage{amsmath,bm}
\usepackage{caption}
\usepackage{subcaption}
\usepackage{subfloat}
\usepackage{bbold}
\usepackage{comment}

\newtheorem{theorem}{\bf Theorem}[section]
\newtheorem{proposition}{\bf Proposition}[section]

\newtheorem{corollary}{\bf Corollary}[section]
\newtheorem{remark}{\bf Remark}[section]
\newtheorem{definition}{\bf Definition}[section]

\newtheorem{example}{\bf Example }[section]

\usepackage{color}

\newenvironment{proof}{
\begin{trivlist}
\item[\hspace{\labelsep}{\bf\noindent Proof. }] }{\par\hfill\end{trivlist}
\par}

\date{\empty}

\title{
\huge\bf Analysis of reliability systems via Gini-type index
\date{Author's version.  Published in:  {\em European Journal of Operational Research}\ 264 (2018), pp.\ 340-353,   
doi:   10.1016/j.ejor.2017.06.013  --
URL: https://www.sciencedirect.com/science/article/pii/S0377221717305313}
}

\author{
\large \bf Motahareh Parsa\footnote{
Department of Statistics, Ordered and Spatial Data Center of Excellence, Ferdowsi University of Mashhad, 
P.O.\ Box 1159, Mashhad 91775, Iran}
\qquad
Antonio Di Crescenzo\footnote{
Corresponding author. 
Dipartimento di Matematica, Universit\`a degli Studi di Salerno, Via Giovanni Paolo II n.132, 84084 Fisciano (SA), Italy, 
E-mail address: adicrescenzo@unisa.it}
\qquad
Hadi Jabbari\footnote{
Department of Statistics, Ordered and Spatial Data Center of Excellence, Ferdowsi University of Mashhad, 
P.O.\ Box 1159, Mashhad 91775, Iran}
}

\begin{document}
 
\maketitle

\begin{abstract}
Different strategies of reliability theory for the analysis of coherent systems 
have been studied by various researchers. Here, the Gini-type index is utilized as an applicable tool 
for the study and  comparison of 
the ageing properties of complex systems. A new stochastic order in terms of Gini-type index is introduced to compare the speed of ageing of components and systems. The parallel-series and series-parallel systems with shared components are studied by their corresponding Gini-type indexes. 
Also, the generalization of Gini-type index for the multidimensional case 
is discussed, and is used to compare components lifetimes properties in the presence of other dependent components. It is shown that the ageing properties of a component lifetime can differ when the other components are working or have already failed. 
Numerous illustrative examples are given for better intuition of Gini-type and 
generalized Gini-type indexes throughout the paper.

\bigskip\noindent
\noindent{\bf Keywords:} 
Reliability, Ageing properties, Stochastic comparisons, 
Systems with shared components, Multivariate conditional hazard rates.

\end{abstract}

\section{Introduction} \label{sec1}
Optimizing the system lifetime is a relevant problem in reliability theory, 
and leads to interesting questions in mathematical statistics and probability modelling. 
Many investigations have been oriented to the development of optimization 
strategies under various assumptions, especially when the system components 
are assumed to be dependent. Spizzichino \cite{Spizzichino 2001} studied notions of dependence 
and notions of ageing, which provide the tools to obtain inequalities for conditional survival probabilities. 
Navarro et al.\ \cite{Navarro 2005} studied some comparisons between coherent systems with dependent components. 
Khaledi and Shaked \cite{Khaledi and Shaked 2007} stochastically compared the residual 
lifetimes of coherent systems with identical or different types of components. 
Navarro et al.\ \cite{Navarro 2013} obtained ordering properties for coherent systems with possibly dependent identically distributed components. Their results are based on a representation of the system reliability function as a distorted function of the common component reliability function. 
\par
Recently, Navarro et al.\ \cite{Navarro 2015} considered a general coherent system with independent or dependent components, and assumed that the components are randomly chosen from two different stocks. They provided sufficient conditions on the components lifetimes and on the random number of components chosen from the two stocks in order to improve the reliability of the whole system. 
See, also, Navarro and Spizzichino 
\cite{NavarroSpizzichino} and Di Crescenzo and Pellerey \cite{DiCrescenzoPellerey} for 
the analysis and the comparison of parallel and series systems with heterogeneous components sharing the same copula, or with components linked via suitable mixtures. 
\par
The stochastic order-based approach has also been exploited by 
Gupta et al.\ \cite{Gupta 2015}, 
aiming to compare the residual lifetime and the inactivity time   
of a used coherent system with the lifetime of the similar coherent system composed by used components. 
A new notion for the comparison of the hazard rates of  random lifetimes has been introduced 
by Belzunce et al.\ \cite{Belzunce 2015}, where the mutual dependence is taken into account.
\par
Borgonovo et al.\ \cite{Borgonovo 2016} studied modern digital systems which may exhibit a non-coherent behaviour and measured the importance of system components. They also proposed a new importance measure for time-independent reliability analysis. 
\par
More recently, Navarro et al.\ \cite{Navarro 2017} provide a general procedure based on the recent concept of generalized distorted distributions to get representations for the reliability functions of inactivity times of coherent systems with dependent components, by which one can compare systems inactivity times.
\par
Here, we propose to adopt new applicable tools to gain 
information on the ageing characteristics of reliability systems, based 
on the Gini-type (GT) index defined by 
Kaminskiy and Krivtsoz \cite{Kaminskiy and Krivtsoz 2010}. 
Such index is expressed in terms of the cumulative hazard rate function of a random lifetime.  
Its definition recalls the well-known `Gini coefficient', which is largely used in the economics literature to analyse incoming distributions. In the context of system reliability, 
the GT index is a measure of the ageing property of a random lifetime. 
As a consequence, in the analysis of point process describing the occurrence of system failure times, the 
GT index is useful to determine if the system is stable, or is improving, or is deteriorating. 
Our aim is to investigate various properties of that index, and to define a new proper stochastic order 
based on it, which is helpful to assess the ageing properties of random lifetimes. 
Specific applications, which involve comparisons and ageing properties of 
parallel-series and series-parallel systems with shared components, are then provided.
\par
Some extensions of GT index to multidimensional case are also thoroughly 
investigated, by which the multivariate and conditional ageing properties of the 
lifetime variables are accessible. It is worth pointing out that our approach allows to study components 
lifetimes properties in the presence of other sharing dependent components. 
Specifically, we show that the multidimensional GT index is able to describe how the 
ageing properties of a component lifetime can vary when the other (dependent) components are working or have already failed.  
 \par
This paper is organized as follows. In Section \ref{sec2} the GT index is introduced and its application in reliability theory is expressed.  
A characterization result of the Weibull distribution in terms of the GT index is also provided. 
In Section \ref{sec3} we define the preannounced stochastic order in terms 
of GT index and discuss its properties. Section \ref{sec4} is devoted to application of the GT index 
to series systems.
In Sections \ref{sec5} and \ref{sec6} two structures for complex systems with shared components 
are introduced. Their ageing properties are studied and compared by means of the GT index. 
\par
In Section \ref{sec7} we define the generalized GT index for bivariate and multivariate random 
lifetimes of the components working in the same environment. 
Furthermore, a new stochastic order is defined
in terms of the generalized GT index and of suitable cumulative hazards. 
Such stochastic order is useful to compare 
the lifetimes of components in the presence of other dependent components and under various 
operational conditions.
\par
Finally, in Section \ref{sec8} the vector GT index is introduced 
for non-negative random variables, based on the multivariate failure rate of multiple components 
of a system. 
\par
Note that throughout this paper,  we say that $X$ is a random lifetime 
in order to refer to a non-negative absolutely continuous (a.c.)\ random variable with   
continuous density function (d.f.). 
Moreover, `$\log$' means natural logarithm, and prime denotes derivative.
\section{Gini-type index} \label{sec2}
The ageing behaviour of repairable or non-repairable systems is vitally important for maintenance strategies. 
Kaminskiy and Krivtsoz \cite{Kaminskiy and Krivtsoz 2010} introduced a simple index which 
helps to assess the degree of ageing, or rejuvenating, 
of repairable (or non-repairable) systems. Let $X$ be a non-negative 
random lifetime of a component or a system. For $t\geq 0$, let
\[
\bar{F}(t)=\mathbb P(X > t)
\]
and
\begin{equation}\label{e1}
H_{\scriptscriptstyle X}(t)=-\log{\bar{F}(t)}
\end{equation}
represent its survival function and cumulative hazard rate function, respectively. 
Assuming that 
\begin{equation}\label{qq1}
 D^1_{\scriptscriptstyle X}:=\{t >0 : \ 0<\bar{F}(t)<1\},
\end{equation}
the GT index is introduced for all $t\in D^1_{\scriptscriptstyle X}$
as follows (see \cite{Kaminskiy and Krivtsoz 2010}). 
We recall that $h_{\scriptscriptstyle X}(t)=\frac{d}{dt}H_{\scriptscriptstyle X}(t)$, 
$t \in D_{\scriptscriptstyle X}^1$, is the hazard rate of $X$. 
\begin{definition} \label{def1} 
The GT index for a random lifetime  $X$, in time interval $ (0,t]$, is
\begin{equation}\label{eq1}
GT_{\scriptscriptstyle X}(t)
=1-\frac{2}{t \,H_{\scriptscriptstyle X}(t)}\int^t_0 H_{\scriptscriptstyle X}(u)\,{\rm d}u, 
\qquad t \in D_{\scriptscriptstyle X}^1.
\end{equation}
\end{definition}
It is shown that GT index satisfies the inequality 
$$ 
 -1 < GT_{\scriptscriptstyle X}(t) < 1 \quad  \hbox{for all $ t \in D_{\scriptscriptstyle X}^1 $.}
$$
Let us now recall some well-known ageing notions that will be related to the properties of GT index.
\begin{definition}\label{defIFR}
A random lifetime $X$ is said to be \\
-- IFR (increasing failure rate)  if $h_{\scriptscriptstyle X}(t)$ is non-decreasing $\forall t$, \\
-- IFRA (increasing failure rate average)  if $H_{\scriptscriptstyle X}(t)/t$ is non-decreasing $\forall t$. \\
Dually, $X$ is said to be \\
-- DFR (decreasing failure rate)  if $h_{\scriptscriptstyle X}(t)$ is non-increasing $\forall t$, \\
-- DFRA (decreasing failure rate average)  if $H_{\scriptscriptstyle X}(t)/t$ is non-increasing $\forall t$. 
\end{definition}
For more details on the aforementioned notions,   
see Shaked and Shanthikumar \cite{Shaked and Shanthikumar 2007} or 
Barlow and Proschan \cite{Barlow-Proshan}.
\par
We point out that the GT index can be assumed as a measure of the ageing property of the 
underlying random lifetime. Indeed, since $H_{\scriptscriptstyle X}(0)=0$ and 
$H_{\scriptscriptstyle X}$ is  an a.c.\ function,  
the following result  holds (see \cite{Kaminskiy and Krivtsoz 2010}). 
\begin{proposition} \label{prop:IFRDFR}
For a  random lifetime $X$ we have that\\
(i)   $GT_{\scriptscriptstyle X}(t) \geq (\leq )\; 0$ for all $t\in D_{\scriptscriptstyle X}^1$ if and only if  $X$ is IFR (DFR);
\\
(ii)   $GT_{\scriptscriptstyle X}(t) =0 $ for all $ t \in D_{\scriptscriptstyle X}^1$ 
if and only if $X$ is CFR (constant failure rate), i.e.\ $X$ has exponential distribution.
\end{proposition}
\par
Clearly, the GT index changes its sign when the hazard rate is 
non-monotonic. For instance, if $h_{\scriptscriptstyle X}(t)= t (t - 1)^2$, $t\geq 0$, then 
$GT_{\scriptscriptstyle X}(t)=\frac{3}{5} + \frac{4 (t-2)}{5 (6 + t (3 t-8)}$, $t\geq 0$, which 
is first positive, then negative, and finally positive as $t$ increases. 
\par
According to \cite{Kaminskiy and Krivtsoz 2010}, it should be mentioned that the GT index is, in a 
sense, distribution-free. 
Moreover, the index introduced in 
Definition \ref{def1} is defined similarly as the `Gini coefficient', which is used in macroeconomics 
for analysing income distributions. In the reliability analysis of repairable systems, it is highly 
interesting to distinguish if the point process of the failure times is close to,  
or far from, the homogeneous 
Poisson process. The analysis of the GT index is thus useful to determine if the system 
is stable, or is improving, or is deteriorating. Indeed, if $X$ describes the consecutive failure times in 
a repairable system, the condition that $GT_{\scriptscriptstyle X}(t)$ is positive (negative) expresses 
that the system is deteriorating (improving), whereas a $GT_{\scriptscriptstyle X}(t)$ vanishing 
means that $X$ has exponential distribution, i.e.\ the system is a homogeneous Poisson process 
(see \cite{Kaminskiy and Krivtsoz 2010} for further details). 
\par
Let us now investigate if the GT index can be a constant value for further 
cases rather than the exponential distribution. This provides us a characterization result 
of the Weibull distribution in terms of the GT index, which extends case (ii) of 
Proposition \ref{prop:IFRDFR}. 
\begin{theorem}
The random lifetime $ X $ has Weibull distribution if and only 
if the corresponding GT index is constant.
\begin{proof}
The proof of one side is straightforward. Thus, suppose that GT index of  $ X $ is constant, 
i.e.\ $GT_{\scriptscriptstyle X}(t)=r$ for $t \in D_{\scriptscriptstyle X}^1$, with $-1<r<1$. 
Hence, from (\ref{eq1}) we have 
$$
\frac{1}{t \,H_{\scriptscriptstyle X}(t)}\int^t_0 H_{\scriptscriptstyle X}(u) \, {\rm d}u
=\frac{1-r}{2},\qquad t \in D_{\scriptscriptstyle X}^1.
$$
Differentiating both sides 
with respect to $t$, since $ H_{\scriptscriptstyle X}(t)$ is differentiable, we obtain
\begin{equation} \label{eqnew3} 
H^{'}_{\scriptscriptstyle X}(t)-\frac{1+r}{1-r}\,\frac{H_{\scriptscriptstyle X}(t)} {t} =0. 
\end{equation}
By solving the differential equation  in \eqref{eqnew3}, 
with $H_{\scriptscriptstyle X}(0)=0$, and using \eqref{e1} one attains  
$$
 \bar{F}(t)=\exp \left\{-c \,t^{\frac{1+r}{1-r}}\right\}, 
 \qquad -1<r<1,\quad  t \in D_{\scriptscriptstyle X}^1,
$$
with $ c >0$, which reveals the survival function of the Weibull distribution. 
\end{proof} 
\end{theorem}
The following example presents a number of distributions, where $D_{\scriptscriptstyle X}^1=(0, \infty)$, 
with the corresponding GT index and the related limit behaviour.

\begin{example}\label{ex1}
Consider the following survival functions, having support $(0,\infty)$: 
\begin{itemize}

\item[(i)]
(Lomax distribution) 
$ \bar{F}(t)=(1+\frac{t}{\beta})^{-\alpha} $, $ \alpha>0 $ and $ \beta >0 $;

\item[(ii)]
(Gompertz Makeham distribution) 
$ \bar{F}(t)=\exp\{\alpha (1-e^{\beta t})\} $, $ \alpha >0 $ and $ \beta>0 $;

\item[(iii)]
(Log-logistic distribution)
$ \bar{F}(t)=\frac{1}{1+(t \beta)^\alpha} $, $ \alpha>0 $ and $ \beta >0 $;

\item[(iv)]
(A bathtub-shaped hazard rate distribution)
$ \bar{F}(t)=\exp\{-\alpha \frac{t^3}{3}+\alpha \beta t^2 -(\alpha\beta^2 +\lambda)t \} $, $ \alpha >0 $, 
$ \lambda >0$ and $ \beta >0 $, having the bathtub-shaped hazard rate $ h_X(t)=\alpha (t-\beta)^2+\lambda $. 
\end{itemize}
Table \ref{tb1} gives the corresponding GT indexes and their 
limits. In case (iii),  $ \Phi$ denotes the Lerch transcendence function, defined as 
\begin{equation}
 \Phi(z,s,a)=\sum^\infty_{k=0} \frac{z^k}{(k+a)^s}.
 \label{eq:Lerch}
\end{equation} 
For a better intuition about the GT indexes in Table \ref{tb1}, see Figures $1$--$4$. 
Note that, due to Proposition \ref{pro2} given in the following section, the GT index for cases (I) and (II) does not depend on $\alpha$.
\begin{table}[t]
\begin{center}
\begin{tabular}{|c|c|c|c|}

\hline 
& $ GT_{\scriptscriptstyle X}(t) $ & $ t\rightarrow 0$ & $ t\rightarrow \infty $\\
\hline
(i) &  $1-2\displaystyle \frac{(\beta +t) \log (1+\frac{\beta}{t})-t }{t \log (1+\frac{\beta}{t})}$ & $0$ & $-1$ \\
\hline
(ii) &  $1+2\displaystyle \frac{e^{\beta t}- \beta t -1}{ \beta t(1-e^{\beta t})}$ & $0$ & $1$ \\
\hline
(iii) &  $1+\displaystyle\frac{2(-(t \beta)^\alpha \Phi(-(t\beta)^\alpha ,1,1+\frac{1}{\alpha})+\log(1+(t \beta)^\alpha))}{-\log(1+(t \beta)^\alpha)}$ &  & $ -1 $ \\
\hline
(iv) & $\displaystyle\frac{ \alpha t^2 - 2 \alpha \beta t}
{ 2 \alpha t^2 - 6 \alpha \beta t + 6 (\alpha \beta ^2 + \lambda) }$ & $0$ & $\displaystyle\frac{1}{2}$ \\
\hline
\end{tabular}
\end{center}
\caption{GT indexes and their limits for the distributions of Example \ref{ex1}.}
\end{table} \label{tb1}
\begin{figure}[ht]
\begin{center}
\includegraphics[scale=0.67]{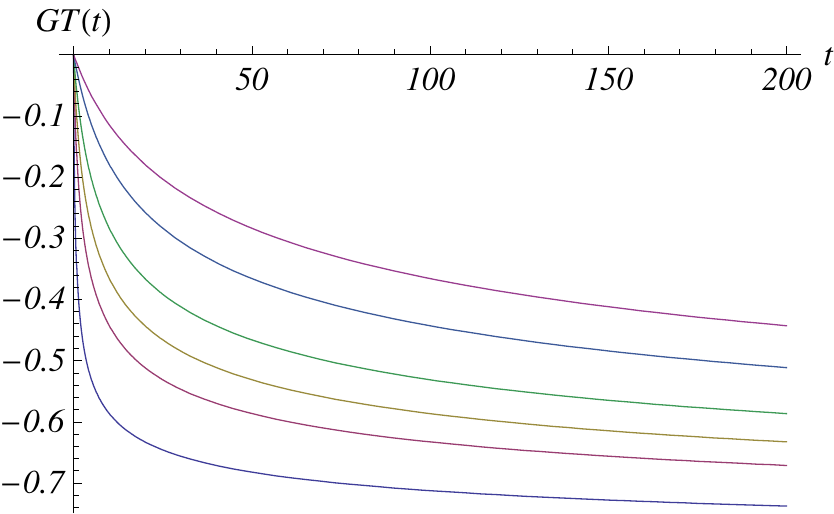}
\caption{The GT index of the Lomax distribution  
(i) for $\beta=0.1, 0.5, 1, 2, 5, 10$ (from bottom to top). }
\end{center}
\end{figure}\label{fig0}
\begin{figure}[ht]
\begin{center}
\includegraphics[scale=0.67]{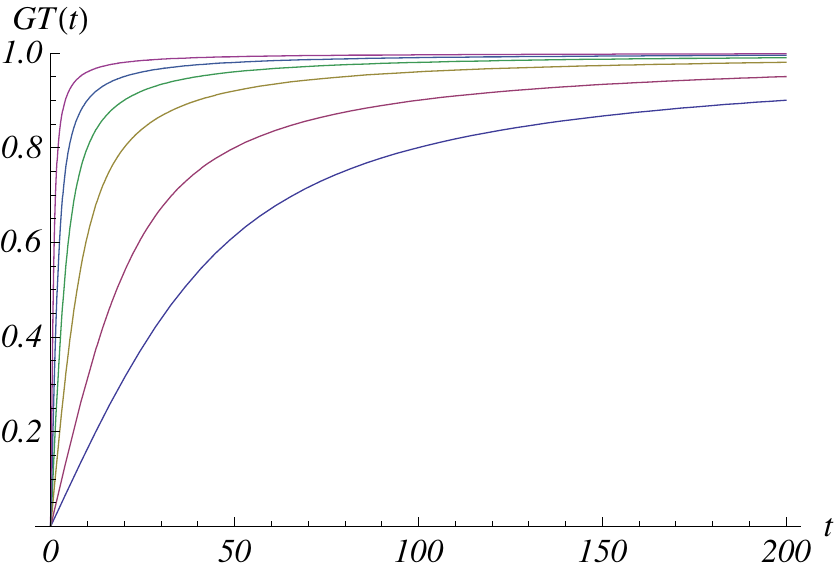}
\caption{The GT index of the Gompertz Makeham distribution 
(ii) for $\beta=0.1, 0.5, 1, 2, 5, 10$ (from bottom to top).}
\end{center}
\end{figure}\label{fig00}
\begin{figure}[ht]
\begin{center}
{\includegraphics[scale=.67]{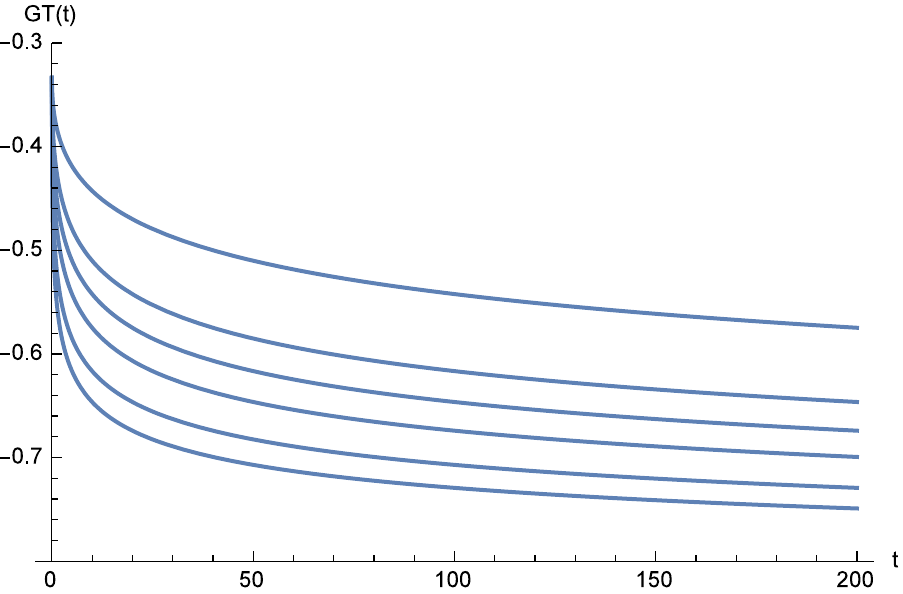}}
{\includegraphics[scale=.67]{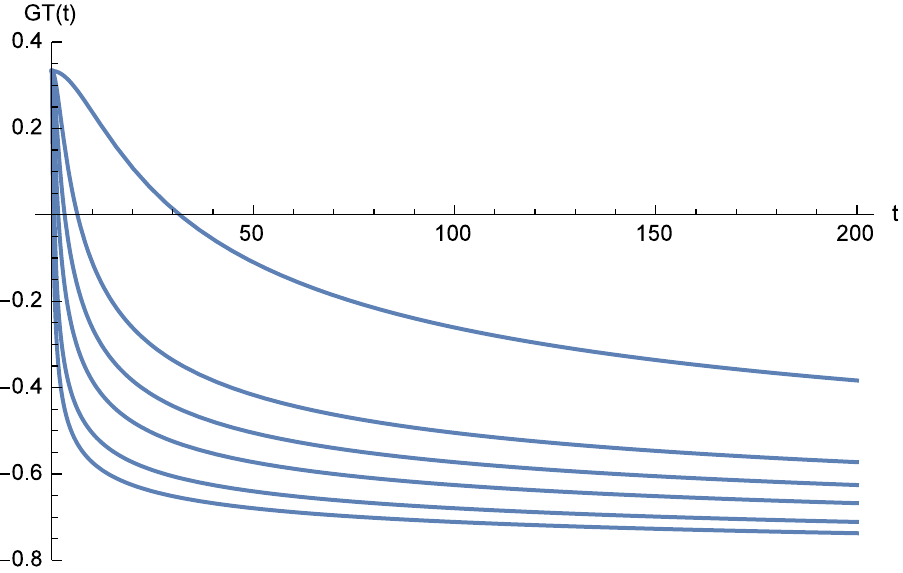}}
\caption{The GT index of the Log-logistic distribution 
(iii) for $\alpha=0.5$ (left) and $ \alpha=2$ (right) where $ \beta=0.1, 0.5, 1, 2, 5, 10$ (from bottom to top).}
\end{center}
\end{figure}\label{fig000}
\begin{figure}[ht]
\begin{center}
{\includegraphics[scale=.67]{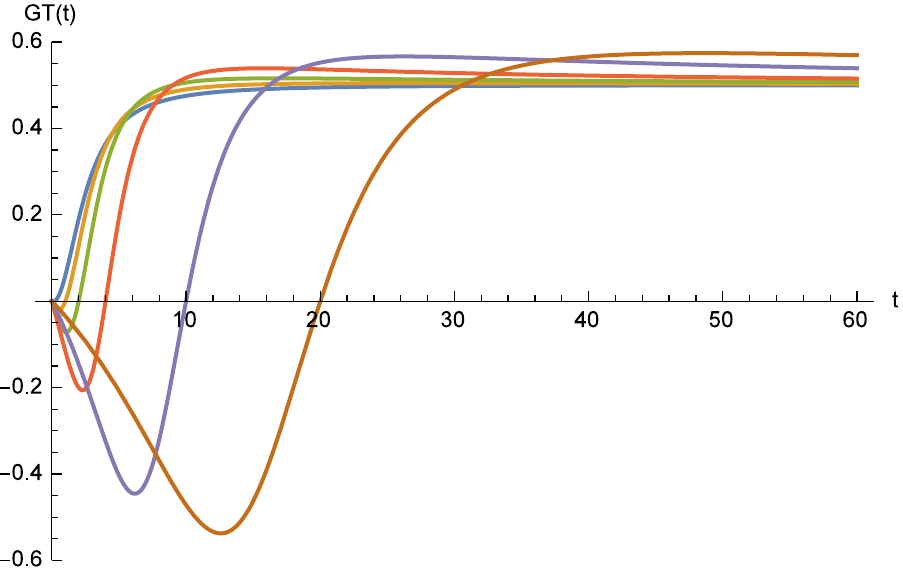}}
{\includegraphics[scale=.67]{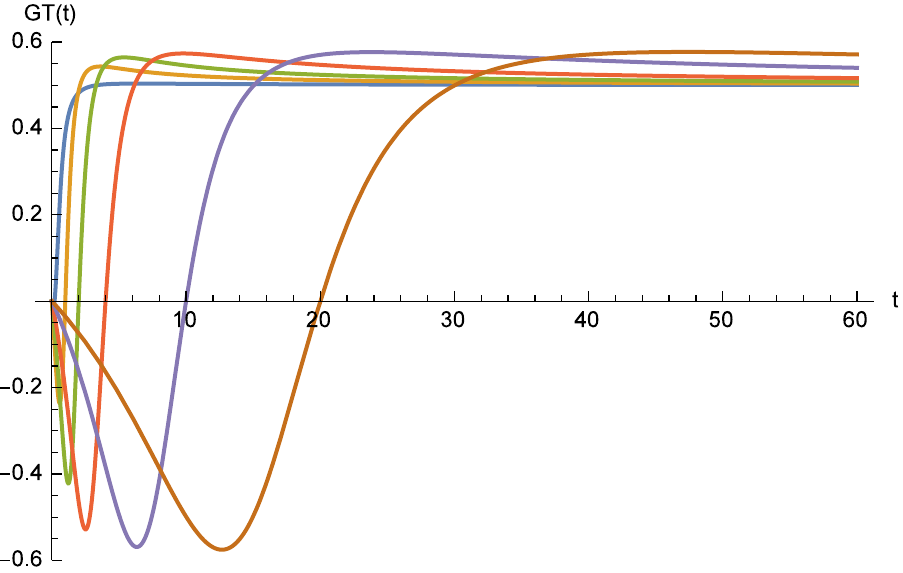}}
\caption{The GT index of the bathtub-shaped hazard rate distribution  
(iv) for $ \alpha=0.5 $ (left) and $ \alpha=10 $ (right) where  $ \lambda=1 $ and $\beta=0.1, 0.5, 1, 2, 5, 10$ 
(from left to right).}
\end{center}
\end{figure}\label{fig0000}
\end{example}
%
\section{A GT index-based stochastic order}\label{sec3}
Stochastic orders are largely employed to compare performances or reliability 
of coherent systems, and to acquire ageing properties of random lifetimes. Let us now define 
a new stochastic order in terms of GT index, called `GT order' for short. 
\begin{definition}\label{df2}
Let $X$ and $ Y $ be two random lifetimes having GT indexes $ GT_{\scriptscriptstyle X} $ and 
$ GT_{\scriptscriptstyle Y}$, respectively. 
We say that $ X $ is less than $ Y $ in GT index, and write $ X \leq_{GT} Y $, if 
$$ 
 GT_{\scriptscriptstyle X}(t) \leq GT_{\scriptscriptstyle Y}(t) 
 \quad \hbox{for all $ t \in D^1:=D^1_{\scriptscriptstyle X} \cap D^1_{\scriptscriptstyle Y}$}.
$$
\end{definition}

An equivalent condition for the GT order is stated hereafter.

\begin{proposition}\label{pro1}
For the random lifetimes $X$ and $ Y $ we  have  $ X \leq_{GT} Y $ if, and only if, 
\begin{equation} \label{eqnew2}
\frac{\int^t_0 H_{\scriptscriptstyle Y}(u) \, {\rm d}u}{\int ^t_0 H_{\scriptscriptstyle X}(u) \, {\rm d}u},
\qquad \hbox{ is non-decreasing in $ t \in D^1 $.}
\end{equation}
\begin{proof}
By Definition \ref{df2} and Eq.\ \eqref{eq1}, one concludes that $ X \leq_{GT} Y $ if,  
and only if, the following condition is fulfilled for all $ t \in D^1$: 
\begin{equation}\label{eqnew}
H_{\scriptscriptstyle Y}(t) \int^t_0 H_{\scriptscriptstyle X}(u) \, {\rm d}u
-H_{\scriptscriptstyle X}(t) \int^t_0 H_{\scriptscriptstyle Y}(u) \, {\rm d}u \geq 0,
\end{equation}
Therefore,  it is not hard to see that inequality (\ref{eqnew}) is equivalent to \\
$ \frac{d}{dt}({\int^t_0 H_{\scriptscriptstyle Y}(u)\, {\rm d}u}/{\int^t_0 H_{\scriptscriptstyle X}(u)\, {\rm d}u}) \geq 0$,  
and then the stated result holds. 
\end{proof}
\end{proposition}
\begin{example}
Let $X(a)$, $a>0$, be a family of  random lifetimes having 
survival functions  $\bar{F}_{\scriptscriptstyle X(a)}(t)=t^a$, $0\leq t\leq 1$. The GT index of 
$X(a)$ is given by 
$$
  GT_{\scriptscriptstyle X(a)}(t)
  =1-\frac{2 }{\log \left(1-t^a\right)}\Big[t^a \Phi \Big(t^a,1,1+\frac{1}{a}\Big)+\log \left(1-t^a\right)\Big],
 \quad 0<t<1,
$$
where $\Phi$ is the Lerch function defined in (\ref{eq:Lerch}). It can be seen that the GT order of $X(a)$ is increasing in $a$, i.e.\ $GT_{\scriptscriptstyle X(a)}(t)<GT_{\scriptscriptstyle X(b)}(t)$ for all 
$a$ and $b$ such that $0<a<b<1$ and all $t\in (0,1)$. 
\end{example}
\par
In the following proposition, we show that the random lifetimes 
$ X $ and $ Y $ that have common support $D^1$ and  
satisfy the proportional hazard rate model
\begin{equation} \label{eqn2}
\bar{F}_{\scriptscriptstyle Y}(t)=\left[\bar{F}_{\scriptscriptstyle X}(t)\right]^{\alpha}, 
\qquad  t\in D^1,~~ \alpha >0,
\end{equation}
if and only if they acquire the same GT indexes. 
\begin{proposition}\label{pro2}
The non-negative random lifetimes $ X $ and $ Y $ which have the 
common support $D^1$ and satisfy the proportional hazard rate property given in \eqref{eqn2}, if and only if
\begin{equation}\label{eqn3}
 GT_{\scriptscriptstyle X}(t)=GT_{\scriptscriptstyle Y}(t),\qquad \forall t \in D^1. 
\end{equation}
\begin{proof}
Let $ H_{\scriptscriptstyle X}(t)$ and $ H_{\scriptscriptstyle Y}(t) $ represent the cumulative hazard 
rate functions for $ X$ and $ Y $, respectively. 
By the proportional hazard rate property \eqref{eqn2} we have $ H_Y(t)=\alpha \,H_X(t) $, $ t\in D^1 $, 
$\alpha>0$. Thus, \eqref{eqn3} holds trivially.
\par
On the other hand, when \eqref{eqn3} holds, according to \eqref{eqnew} one concludes that 
$  {\int^t_0 H_{\scriptscriptstyle Y}(u)\,{\rm d}u}/{\int^t_0 H_{\scriptscriptstyle X}(u)\,{\rm d}u}=\alpha $, 
where $ \alpha >0 $, which implies the proportional hazard rate property. 
\end{proof}
\end{proposition}
\begin{remark}
Due to Proposition \ref{pro2}, the antisymmetry property of GT order does not hold in a  
strict sense. Indeed, $ X \leq_{GT} Y $ and $ Y \leq_{GT} X $ are achieved simultaneously 
if, and only if, $ X $ and $ Y $ satisfy the proportional hazard rate model.
\end{remark}
\par
Let us now discuss some further properties of the GT order.
\begin{proposition}\label{pro3}
Let $ X $, $ Y $ and $ Z $ be random lifetimes. The following properties hold:
\begin{itemize}
\item[(i)] (reflexivity) $ X \leq_{GT} X $.

\item[(ii)] (transitivity) If $ X \leq_{GT} Y $ and $ Y \leq_{GT} Z$ then $ X \leq_{GT} Z $.

\item[(iii)] $X \leq_{GT} Y$ \quad $\Longleftrightarrow$  \quad $aX+b \leq_{GT} aY+b$ 
for $ a, b \in \mathbb R^{+} $. 
\end{itemize}
\begin{proof}
The proof is trivial by Proposition \ref{pro1}.
\end{proof}
\end{proposition}
Sengupta and Deshpande \cite{Sengupta and Deshpande 1994}  introduced the 
following partial orderings dealing with ageing properties of random lifetimes. 
Hereafter, we slightly modify their definitions in order to have more consistent notions.
Recall the concepts given in Definition \ref{defIFR}. 
\begin{definition}\label{def4}
Given the random lifetimes $ X $ and $ Y $, we say that $ X $ is: \\
-- ageing slower than $ Y $, and write $ X \leq_c Y $, if 
$h_{\scriptscriptstyle Y}(t)/h_{\scriptscriptstyle X}(t)$ is non-decreasing in $ t \in D^1 $, or 
equivalently if $Z=H_{\scriptscriptstyle Y}(X)$ is IFR;
\\
-- ageing slower than $ Y $ in average, and write $ X \leq_\star Y $, 
if $H_{\scriptscriptstyle Y}(t)/H_{\scriptscriptstyle X}(t)$ is non-decreasing in $ t \in D^1 $, or 
equivalently if $Z=H_{\scriptscriptstyle Y}(X)$ is IFRA.
\end{definition}
And therefore we have 
\[
 X \leq_c Y \quad \Longrightarrow  \quad X \leq_\star Y.
\]
We stress that the inequalities  given in Definition \ref{def4}, have been inverted with respect 
to the corresponding orderings  given in \cite{Sengupta and Deshpande 1994}. 
\par
The following proposition expresses the relation between the GT order and the aforementioned 
lifetime orders.
\begin{proposition}\label{pro5}
Suppose that at least one of the  random lifetimes $ X $ and $ Y $ has a 
strictly increasing distribution. Then we have 
\begin{equation}\label{n111}
 X \leq_\star Y \quad \Longrightarrow  \quad X \leq_{GT} Y.
\end{equation}
Moreover, if $\psi$ is a strictly increasing positive function passing through $ (0,0) $ then  
\begin{equation}\label{n222}
X \leq_\star Y \quad\Longrightarrow \quad\psi(X) \leq_{GT} \psi(Y).
\end{equation}
\begin{proof}
According to Proposition 2.3 in \cite{Sengupta and Deshpande 1994}, we have $X \leq_\star Y$ 
if and only if $ H_{\scriptscriptstyle Y}(t)/ H_{\scriptscriptstyle X}(t) $ is non-decreasing in $ t \in D^1$. 
Since $ H_{\scriptscriptstyle X}(u)$ and $ H_{\scriptscriptstyle Y}(u)$ are non-negative non-decreasing 
functions of $ u$, then it is not hard to see that the condition \eqref{eqnew2} in Proposition \ref{pro1} 
is held, so that $X \leq_{GT} Y$. 
\par
Theorem 2.1 in \cite{Sengupta and Deshpande 1994} states that $X \leq_\star Y$ 
if and only if $ \psi(X) \leq \psi(Y) $ for every strictly increasing positive function $\psi$ passing 
through $(0,0)$. Hence, the validity of (\ref{n222}) follows straightforwardly from (\ref{n111}).
\end{proof}
\end{proposition}
\par
The  improvement of  the reliability of  coherent systems is often obtained by adding redundance or 
by performing replacement of components 
(e.g., see the recent contribution by Eryilmaz \cite{Eryilmaz} on $\delta$-shock models.) 
The stochastic ordering introduced in Definition \ref{df2} is useful to achieve the ageing properties of   
random lifetimes. Indeed, due to the results above,  condition $X \leq_{GT} Y$ 
means that $X$ is ageing slower than $Y$ in a broad sense. Hereafter, as an example,  
we analyse the problem of improving a coherent system by adding 
a redundant component, and compare its properties in terms of the GT index. 
\begin{example}\label{ex:newex}
Consider the series-parallel coherent system shown in Figure 5 (see Example 1 of 
Doostparast et al.\ \cite{Doostparastetal}), whose 11 components have i.i.d.\ random lifetimes $X_i$'s. 
Assume that the system can be improved by including 
redundancy, specifically by changing the component having lifetime $X_6$ into a 2-component parallel subsystem formed by i.i.d.\ components of the same type. Figure 6 shows the GT index 
of the considered systems for the 2 cases when all random lifetimes have the same d.f.:
\\
(a) $F(t)= 1-e^{-t}$, $t\geq 0$ (exponential distribution);
\\
(b)  $F(t)= \frac{t}{t+1}$, $t\geq 0$ (Lomax distribution). 
\\
It is evident that in both cases the parallel redundancy on the component increases 
the GT index of the system lifetime. 
Moreover, Figure 6 indicates that in all cases the GT index of the considered series-parallel system 
starts from $0.5$ and it is decreasing in time, but with different limits. 
When the components lifetimes have exponential distribution, i.e.\ in case (a), the GT indexes 
tend to 0 as $t\to \infty$, whereas, for the Lomax distribution, in case (b), the limits are negative.  
Hence, in case (a) the system lifetime is always IFR though the deteriorating property of this system is reduced 
in time. In case (b), the hazard rate of the system lifetime is non-monotonic; the system is IFR in the beginning but the ageing property changes into DFR after some time. 
This example indicates that the optimal structure and time for the redundancy in complex systems can be determined by assuming a threshold of GT index which specifies the system ageing property. 
\end{example}
 \begin{figure}[t] 
 \begin{center} 
 \begin{picture}(340,170) 
 \put(0,100){\circle*{4}} 
 \put(0,100){\line(1,0){40}}
 \put(20,60){\line(0,1){80}}
 \put(20,60){\line(1,0){20}}
 \put(20,140){\line(1,0){20}}
 \put(40,45){\line(0,1){30}}
 \put(40,85){\line(0,1){30}}
 \put(40,125){\line(0,1){30}}
 \put(40,45){\line(1,0){30}}
 \put(40,75){\line(1,0){30}}
 \put(40,85){\line(1,0){30}}
 \put(40,115){\line(1,0){30}}
 \put(40,125){\line(1,0){30}}
 \put(40,155){\line(1,0){30}}
 \put(70,45){\line(0,1){30}}
 \put(70,85){\line(0,1){30}}
 \put(70,125){\line(0,1){30}}
 \put(50,140){\makebox(10,5)[t]{\footnotesize $X_1$}}
 \put(50,100){\makebox(10,5)[t]{\footnotesize $X_2$}} 
 \put(50,60){\makebox(10,5)[t]{\footnotesize $X_3$}} 
 \put(70,60){\line(1,0){20}}
 \put(70,140){\line(1,0){20}}
 \put(70,100){\line(1,0){85}}  
 \put(90,60){\line(0,1){80}}
 \put(110,60){\line(0,1){80}}
 \put(110,140){\line(1,0){20}}
 \put(130,125){\line(0,1){30}}
 \put(130,155){\line(1,0){30}}
 \put(130,125){\line(1,0){30}}
 \put(180,155){\line(1,0){30}}
 \put(180,125){\line(1,0){30}}
 \put(160,125){\line(0,1){30}}
 \put(180,125){\line(0,1){30}}
 \put(160,140){\line(1,0){20}}
 \put(210,125){\line(0,1){30}}
 \put(210,140){\line(1,0){20}}
 \put(110,60){\line(1,0){45}}
 \put(155,45){\line(0,1){30}}
 \put(155,85){\line(0,1){30}}
 \put(155,45){\line(1,0){20}}
 \put(155,45){\line(1,0){30}}
 \put(155,75){\line(1,0){30}}
 \put(155,85){\line(1,0){30}}
 \put(155,115){\line(1,0){30}}
 \put(185,45){\line(0,1){30}}
 \put(185,85){\line(0,1){30}}
 \put(185,60){\line(1,0){45}}
 \put(185,100){\line(1,0){65}}
 \put(230,60){\line(0,1){80}}
 \put(140,140){\makebox(10,5)[t]{\footnotesize $X_4$}}
 \put(190,140){\makebox(10,5)[t]{\footnotesize $X_5$}}
 \put(165,100){\makebox(10,5)[t]{\footnotesize $X_6$}} 
 \put(165,60){\makebox(10,5)[t]{\footnotesize $X_7$}} 
 \put(250,40){\line(0,1){120}}
 \put(250,40){\line(1,0){20}}
 \put(250,80){\line(1,0){20}}
 \put(250,120){\line(1,0){20}}
 \put(250,160){\line(1,0){20}}
 \put(270,25){\line(0,1){30}}
 \put(270,65){\line(0,1){30}}
 \put(270,105){\line(0,1){30}}
 \put(270,145){\line(0,1){30}}
 \put(270,25){\line(1,0){30}}
 \put(270,55){\line(1,0){30}}
 \put(270,65){\line(1,0){30}}
 \put(270,95){\line(1,0){30}}
 \put(270,105){\line(1,0){30}}
 \put(270,135){\line(1,0){30}}
 \put(270,145){\line(1,0){30}}
 \put(270,175){\line(1,0){30}}
 \put(300,25){\line(0,1){30}}
 \put(300,65){\line(0,1){30}}
 \put(300,105){\line(0,1){30}}
 \put(300,145){\line(0,1){30}}
 \put(300,40){\line(1,0){20}}
 \put(300,80){\line(1,0){20}}
 \put(300,120){\line(1,0){20}}
 \put(300,160){\line(1,0){20}}
 \put(320,40){\line(0,1){120}}
 \put(320,100){\line(1,0){20}}
 \put(340,100){\circle*{4}} 
 \put(280,160){\makebox(10,5)[t]{\footnotesize $X_8$}}
 \put(280,120){\makebox(10,5)[t]{\footnotesize $X_9$}}
 \put(280,80){\makebox(10,5)[t]{\footnotesize $X_{10}$}} 
 \put(280,40){\makebox(10,5)[t]{\footnotesize $X_{11}$}} 
\end{picture} 
\end{center}
   \caption{
   Schematic representation of the  system considered in Example \ref{ex:newex}.
   }
\label{Fig:NewSystem} 
\end{figure}
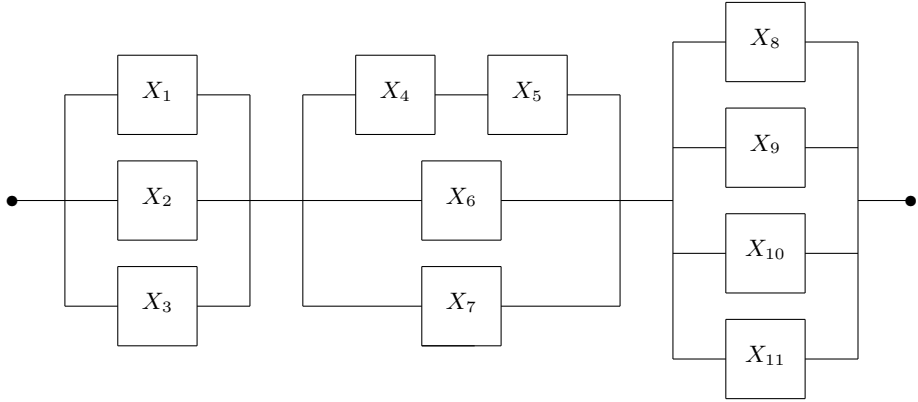
%
\begin{figure}[t]
\begin{center}
{\includegraphics[scale=0.65]{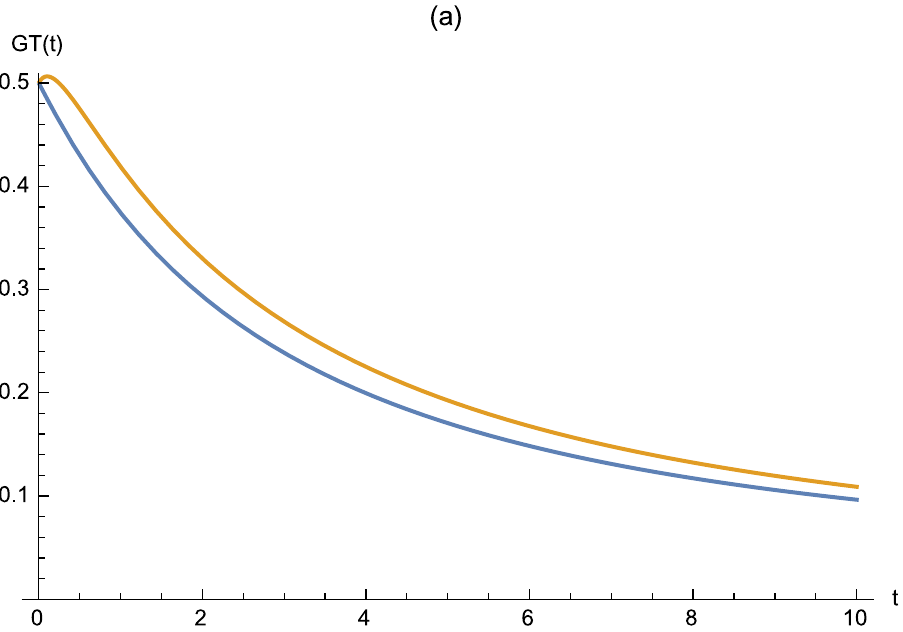}}
{\includegraphics[scale=0.65]{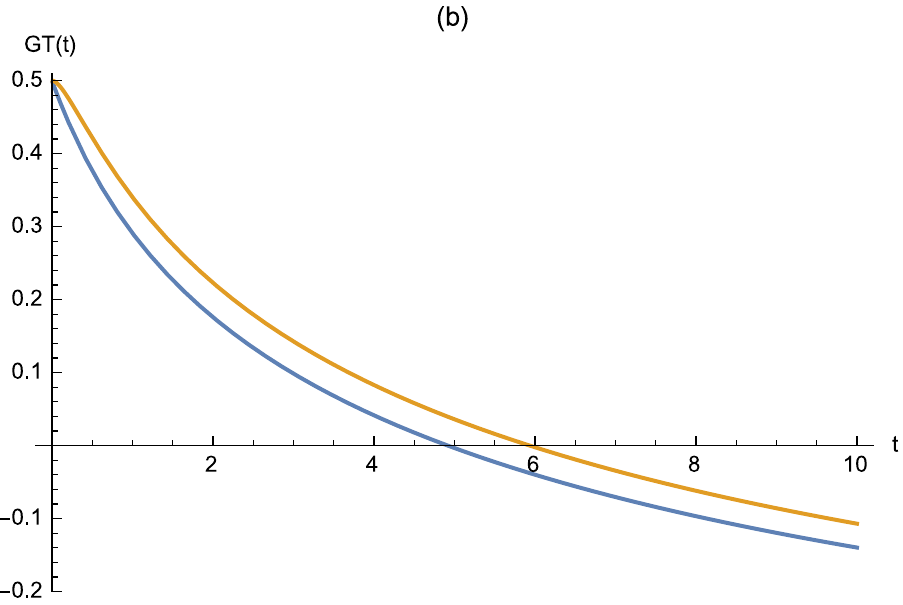}}
\caption{
The GT index for the system of Figure \ref{Fig:NewSystem} (lower curve) and for the redundant 
system (upper curve) of Example \ref{ex:newex} when the lifetimes  have (a) exponential distributions, 
and (b) Lomax distributions.
}
\end{center}
\label{fignew1}
\end{figure}
\par
Example \ref{ex:newex} suggests to propose a redundancy policy for reliability systems 
based on the GT index. Assume that one or more components of the system can be duplicated by 
insertion of a redundant component. This policy is based on the determination of the GT index for all the 
enlarged systems. We recall that a positive (negative) ageing property for random lifetimes 
corresponds to a positive (negative) GT index (cf.\ Proposition  \ref{prop:IFRDFR}). Hence, if the specific application asks for a system lifetime having a positive (negative) ageing property, then the 
preventive assignment of redundant component(s) is performed according to the choice leading to the greatest (lowest) GT index for the time interval of interest. 
Other suitable procedures can be investigated according to the redundancy assignments, in order to 
attain a preassigned threshold level for the system GT index. 
In addition, similarly as in Doostparast et al.\ \cite{Doostparastetal}, the following rule can be implemented:
If the system reliability in terms of GT index reaches a lower threshold, then a unit replacement/repair must occur. 
The detailed analysis of such criteria, as well as setting of the replacement/repair procedures while leading to specific levels of GT indexes, will be the object of a subsequent investigation.
%
\section{Gini-type index for series systems}\label{sec4}
In this section we investigate the GT index of the lifetime of a  
series system composed of $n$ independent components. 
We denote by $X_i$ the lifetime of the $i$-th component, for $i=1,2,\ldots,n$. Let 
\begin{equation}
 T_S=\min_{i\in \{1,2,\ldots,n\}}X_i
 \label{eq:defTS}
\end{equation}
represent the lifetime of the series system, and let
$$
 T_{S\setminus \{j\}}=\min_{i\in \{1,2,\ldots,n\}\setminus \{j\}}X_i
$$
be the lifetime of the series system deprivated of the $j$-th component. 
By making use of the GT order introduced in Definition \ref{df2}, 
the next proposition reveals that the ageing of a series system is larger  
than the ageing of one of its components, if the ageing of the system without that specific  
component is still larger  than the ageing of such unit.
\begin{proposition} \label{pro7}
For the lifetime of a series system composed of $n$ independent components, 
for any  $j=1,2,\ldots,n$ we have
\[ 
T_S \geq_{GT} X_j
 \]
 if and only if 
 \[
 T_{S\setminus \{j\}} \geq_{GT} X_j.
\]
\begin{proof}
Let $ \bar{F}_{X_i}(t)$, for $ t \in D^1_i $, be the survival function of $X_i$, where 
$D^1_i=\{t >0 :0<\bar{F}_{X_i}(t)<1\}$. Similarly, let $ \bar{F}_{T_S}(t)$, for $t\in D^1$, 
be the survival function of $T_S$, where $ D^1=\bigcap^{n}_{i=1} D^1_i$. 
Hence, due to (\ref{eq:defTS}), the cumulative hazard rate of $T_S$, for $t\in D^1$,  is 
\begin{equation*}
H_{T_S}(t)=-\log{\bar{F}_{T_{S}}(t)}=-\sum^n_{i=1}\log{\bar{F}_{X_i}(t)}
=\sum^n_{i=1} H_{X_i}(t).
\end{equation*}
Following the concept introduced in \eqref{eq1}, 
the GT index for the random  lifetime $T_S$ is 
\begin{equation*}
GT_{T_{S}}(t)=1-\dfrac{2\int^t_0 \sum^n_{i=1} H_{X_i}(u)du}{t \sum^n_{i=1} H_{X_i}(t)}, 
\qquad t\in D^1.
\end{equation*}
Assuming that $ GT_{T_{S}}(t) \geq GT_{T_{S\setminus \{j\}}}(t) $, for $ j=1,2,\dots, n $, then 
\begin{equation*}
\dfrac{2\int^t_0 \sum^n_{i=1} H_{X_i}(u)du}{t \sum^n_{i=1} H_{X_i}(t)} \leq \dfrac{2\int^t_0  H_{X_i}(u)du}{t H_{X_i}(t)}.
\end{equation*}
Hence, we obtain 
\begin{equation*}
 H_{X_j}(t)\int^t_0 \sum^n_{i=1} H_{X_i}(u)du \leq t \sum^n_{i=1} H_{X_i}(t)\int^t_0  H_{X_j}(u)du,
\end{equation*}
by which it is concluded that $ GT_{T_{S\setminus \{j\}}}(t) \geq GT_{X_j}(t) $, for $ j=1,2, \dots, n $. 
The reverse can be easily gained using the same method.
\end{proof}
\end{proposition}
From Proposition \ref{pro7} we have that a series system 
is larger than a generic component in the GT index if, and only if, 
the series system without such unit satisfies the same condition. 
This is in agreement with fact that  the series system always deteriorates faster than its single units. 
According to applicable perspective, it might be necessary to estimate the preventive repair 
or replacement strategies, or utilize some redundancy policies such as installing standby units. 
\par
The following results are immediately concluded from Proposition \ref{pro7}.
\begin{corollary}
Under the assumptions of Proposition \ref{pro7}, we have
\begin{itemize}
\item If the independent random lifetimes $ X_i $'s are exponentially distributed, then $GT_{T_S}(t) \geq 0$, if and only if $ GT_{T_{S\setminus\{j\}}}(t) \geq 0$, for $j=1, 2, \dots, n$. This states that the series system is IFR though 
all the components are CFR.

\item If $ n=2 $, then $T_S \geq_{GT} X_1$, if and only if $X_2  \geq_{GT} X_1$.  
Therefore, the series system made of two components is ageing faster than one of its specific components if and only if the other component is ageing faster than the specific one. 
\end{itemize}
\end{corollary}

\par
The following sections consider two applications of special reliability systems having shared components. 
Specifically, it will be shown that the GT index is useful to 
analyse the ageing properties of complex systems. 
\section{Parallel-series system with shared components}\label{sec5}
Let $ X_1,\dots, X_n $ be non-negative independent and identically distributed (iid) random 
variables, which denote the lifetime of $ n $ components working in the same system, having 
cumulative distribution function (c.d.f.) $F$ and survival function $\bar F=1-F$. 
In parallel-series system with shared components we suppose that, for a fixed $k$, $ 1 \leq k < n$,
each $ k $ components are working as a local series system. Thus we set 
\[ 
Y_j=\min_{j \leq i \leq j+k-1} X_i,\qquad j=1,2, \dots, n-k+1.  
\]
This relation defines a local dependence among the components. 
The local series systems are connected in parallel and constitute 
the main system with the lifetime given by 
\begin{equation}
 T_{P-S}=\max_{1 \leq j \leq n-k+1}Y_j. 
 \label{eq:TPS}
\end{equation}
We aim to specify a general rule for determining the c.d.f.\ of $T_{P-S}$, and 
then to evaluate the corresponding GT index in order to assess the ageing property of the system. 
Due to (\ref{eq:TPS}), the c.d.f.\ of the system lifetime is obtained as follows:
\begin{equation}\label{eq2}
\begin{aligned}
G_{n-k+1}(t)&:=\mathbb P\left(T_{P-S}\leq t \right)
=\mathbb P\left(Y_1\leq t, \dots, Y_n \leq t\right).
\end{aligned}
\end{equation}
Conditioning on the failure of the first $ k $ components we get
\begin{equation}\label{eq3}
\begin{aligned}
&G_{n-k+1}(t)=\mathbb P\left[\min_{2 \leq i \leq k+1} X_i \leq t, \dots, \min_{n-k \leq i \leq n} X_i \leq t \right] F(t) \\
&\quad +\mathbb P\left[\min_{3 \leq i \leq k+1}X_i \leq t, \dots, \min_{n-k \leq i \leq n} X_i \leq t \right]F(t)\bar{F}(t) \\
&\quad+\dots \\
&\quad+\mathbb P\left[\min_{k+1 \leq i \leq 2k+1} X_i \leq t, \dots, \min_{n-k \leq i \leq n} X_i \leq t \right]F(t) \bar{F}^{k-1}(t)\\
&\quad+\mathbb P\left[ \min_{1 \leq i \leq k} X_i \leq t, \min_{2 \leq i \leq k+1} X_i \leq t, \dots, \min_{n-k \leq i \leq n} X_i \leq t  \ \bigg| \  X_1 >t ,\dots ,X_k >t \right] \bar{F}^k(t),
\end{aligned}
\end{equation}
in which the last probability is vanishing. 
Thus, the c.d.f.\ of the system lifetime is 
\begin{equation}\label{eq4}
G_{n-k+1}(t)=F(t) \sum^{k}_{i=1} \bar{F}^{i-1}(t) G_{n-k+1-i}(t),\qquad  t \in D_X^1,
\end{equation}
with $D_X^1$ defined in \eqref{qq1}. 
To find the c.d.f.\ $ G_{n-k+1}(t) $ one needs to solve the difference equation of order $ k $ given in \eqref{eq4}, which in general requires numerical methods. 
\par
Formally, the GT index for the considered parallel-series system with shared components is expressed as follows 
\begin{equation}\label{eq5}
GT_{P-S}(t)=1-\frac{2 \int^t_0 \log(1-G_{n-k+1}(u))du}{t \log(1-G_{n-k+1}(t))},\qquad  t \in D_X^1.
\end{equation}
%
\subsection{Case $ k=2 $}
Let us now consider the parallel-series system studied above in the special case $k=2$, 
which can be represented as in Figure 7. When $k=2$ and $n\geq 3$, $n\in \mathbb N$, 
due to \eqref{eq4}, the c.d.f.\ of the system lifetime can be expressed as
\begin{equation}\label{eq6}
G_{n-1}(t)=F(t) G_{n-2}(t)+F(t)\bar{F}(t) G_{n-3}(t), 
\qquad t \in D_X^1.
\end{equation}
To solve \eqref{eq6}, we note that the auxiliary equation
\begin{equation} \label{eq7}
\alpha ^{n-1}(t)-F(t)\alpha ^{n-2}(t)-F(t)\bar{F}(t)\alpha ^{n-3}(t)=0, 
\end{equation}
has the following solution
\begin{equation}\label{eq8}
 G_n(t)=[1-C(t)]\alpha^n_{1}(t)+C(t)\alpha^n_{2}(t),
\end{equation}
where 
$$ 
 \alpha_{1,2}(t)=\frac{1}{2}\left[F(t)\pm \sqrt{F^2(t)+4F(t)\bar{F}(t)}\right], 
 \qquad \alpha_2(t) < 0 < \alpha_1(t),
$$ 
with   
$$ 
 C(t)=\frac{\alpha_1(t) +\bar{F}^2(t)-1}{\alpha_1(t)-\alpha_2(t)}.
$$
 \begin{figure}[t] 
 \begin{center} 
 \begin{picture}(240,270) 
 \put(70,250){\makebox(20,15)[t]{\large $X_1$}}
 \put(70,200){\makebox(20,15)[t]{\large $X_3$}} 
 \put(70,150){\makebox(20,15)[t]{\large $X_5$}}
 \put(70,70){\makebox(20,15)[t]{\large $X_{n-3}$}}
 \put(70,20){\makebox(20,15)[t]{\large $X_{n-1}$}}
 \put(160,250){\makebox(20,15)[t]{\large $X_2$}}
 \put(160,200){\makebox(20,15)[t]{\large $X_4$}} \put(160,150){\makebox(20,15)[t]{\large $X_6$}} \put(160,70){\makebox(20,15)[t]{\large $X_{n-2}$}}
 \put(160,20){\makebox(20,15)[t]{\large $X_{n}$}}
 \put(55,240){\line(0,1){40}} 
 \put(55,190){\line(0,1){40}} \put(55,140){\line(0,1){40}} \put(55,60){\line(0,1){40}}
 \put(55,10){\line(0,1){40}}
 \put(105,240){\line(0,1){40}}
 \put(105,190){\line(0,1){40}} \put(105,140){\line(0,1){40}}
 \put(105,60){\line(0,1){40}}
 \put(105,10){\line(0,1){40}}
  \put(55,280){\line(1,0){50}} \put(55,240){\line(1,0){50}} \put(55,230){\line(1,0){50}}
  \put(55,190){\line(1,0){50}}
  \put(55,180){\line(1,0){50}}
  \put(55,140){\line(1,0){50}}
  \put(55,100){\line(1,0){50}}
  \put(55,60){\line(1,0){50}}
  \put(55,50){\line(1,0){50}}
  \put(55,10){\line(1,0){50}}
   \put(145,240){\line(0,1){40}}
   \put(145,190){\line(0,1){40}}
   \put(145,140){\line(0,1){40}} \put(145,60){\line(0,1){40}}
   \put(145,10){\line(0,1){40}} \put(195,240){\line(0,1){40}}
   \put(195,190){\line(0,1){40}} \put(195,140){\line(0,1){40}}
   \put(195,60){\line(0,1){40}} \put(195,10){\line(0,1){40}} 
   \put(145,280){\line(1,0){50}} \put(145,240){\line(1,0){50}}
   \put(145,230){\line(1,0){50}} \put(145,190){\line(1,0){50}}
   \put(145,180){\line(1,0){50}} \put(145,140){\line(1,0){50}}
   \put(145,100){\line(1,0){50}} \put(145,60){\line(1,0){50}}
   \put(145,50){\line(1,0){50}} \put(145,10){\line(1,0){50}}
   \put(240,145){\circle*{4}} \put(10,145){\circle*{4}} 
   \put(10,145){\line(1,0){20}} \put(220,145){\line(1,0){20}} 
   \put(30,30){\line(0,1){230}} \put(220,30){\line(0,1){230}}
   \put(30,30){\vector(1,0){15}} \put(30,80){\vector(1,0){15}}
   \put(30,160){\vector(1,0){15}} \put(30,210){\vector(1,0){15}}
   \put(30,260){\vector(1,0){15}} \put(45,30){\line(1,0){10}}
   \put(45,80){\line(1,0){10}} \put(45,160){\line(1,0){10}}
   \put(45,210){\line(1,0){10}} \put(45,260){\line(1,0){10}}
   \put(195,30){\vector(1,0){15}} \put(195,80){\vector(1,0){15}}
   \put(195,160){\vector(1,0){15}} \put(195,210){\vector(1,0){15}}
   \put(195,260){\vector(1,0){15}} \put(210,30){\line(1,0){10}}
   \put(210,80){\line(1,0){10}} \put(210,160){\line(1,0){10}}
   \put(210,210){\line(1,0){10}} \put(210,260){\line(1,0){10}}
   \put(80,131){\line(0,1){4}} \put(80,122){\line(0,1){4}}
   \put(80,113){\line(0,1){4}} \put(80,104){\line(0,1){4}}
   \put(170,131){\line(0,1){4}} \put(170,122){\line(0,1){4}}
   \put(170,113){\line(0,1){4}} \put(170,104){\line(0,1){4}} 
   \put(105,30){\vector(1,0){25}} \put(105,80){\vector(1,0){25}}
   \put(105,160){\vector(1,0){25}} \put(105,210){\vector(1,0){25}}
   \put(105,260){\vector(1,0){25}} \put(130,30){\line(1,0){15}}
   \put(130,80){\line(1,0){15}} \put(130,160){\line(1,0){15}}
   \put(130,210){\line(1,0){15}} \put(130,260){\line(1,0){15}}
   \put(105,220){\vector(4,3){25}} \put(105,170){\vector(4,3){25}}
   \put(105,40){\vector(4,3){25}} \put(130,238.75){\line(4,3){15}} \put(130,188.75){\line(4,3){15}} \put(130,58.75){\line(4,3){15}}
   \end{picture} 
   \end{center}
   \caption{Schematic representation of the parallel-series system with shared components when $ k=2 $ and $ n $ is even.}
\label{Fig:System} 
\end{figure}
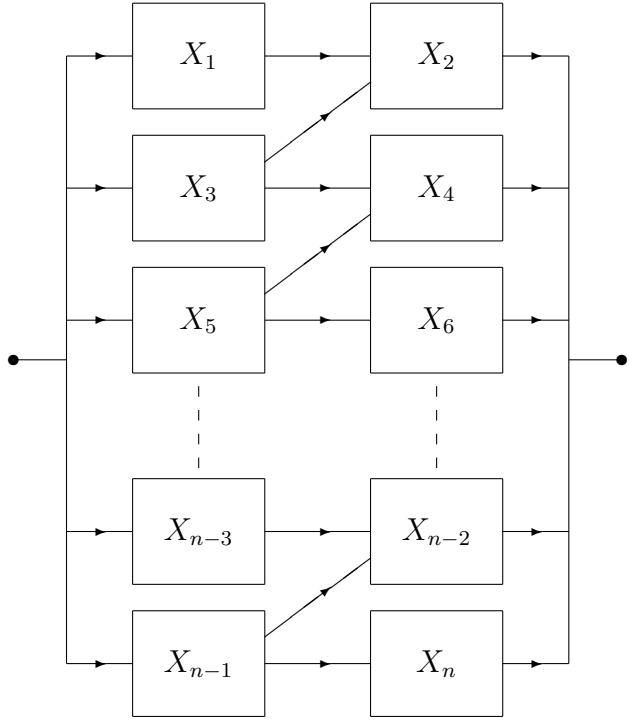
\begin{example}\label{Exm1}
If the i.i.d. random lifetimes $  X_i $'s have Weibull c.d.f.\ $ F(t)=1-e^{-(\lambda t)^m}$ for  $ t>0 $, where $ \lambda>0$ and $m >0 $, then the c.d.f.\ (\ref{eq8}) becomes 
\begin{equation}\label{eq9}
\begin{aligned}
G_{n-1}(t)&=\frac{1}{A(t)}\left[\frac{1}{2}-e^{-2(\lambda t)^m}+\frac{1}{2}e^{-(\lambda t)^m}+\frac{1}{2}A(t)\right] \\
&\times \frac{1}{2^n}\left[1- e^{-(\lambda t)^m}+ A(t)\right]^n \\
&+\left\{1-\frac{1}{A(t)}\left[\frac{1}{2}-e^{-2(\lambda t)^m}+\frac{1}{2}e^{-(\lambda t)^m}+\frac{1}{2}A(t)\right]\right\}\\ 
&\times \frac{1}{2^n}\left[1- e^{-(\lambda t)^m}- A(t)\right]^n, \qquad t>0,
\end{aligned}
\end{equation}
where
\[
A(t)=\sqrt{(1-e^{- (\lambda t)^m})^2+4 (1-e^{-(\lambda t)^m}) e^{(\lambda t)^m}}.
\]
Clearly, the GT index can be determined by means of (\ref{eq5}) and (\ref{eq9}). 
When the random lifetimes have Weibull c.d.f., some values of such GT index 
are illustrated in Figure 8. Also, the behaviour of the GT index versus the 
number of components is shown in Figure 9.
\begin{figure}[ht]
\begin{center}
{\includegraphics[scale=0.3]{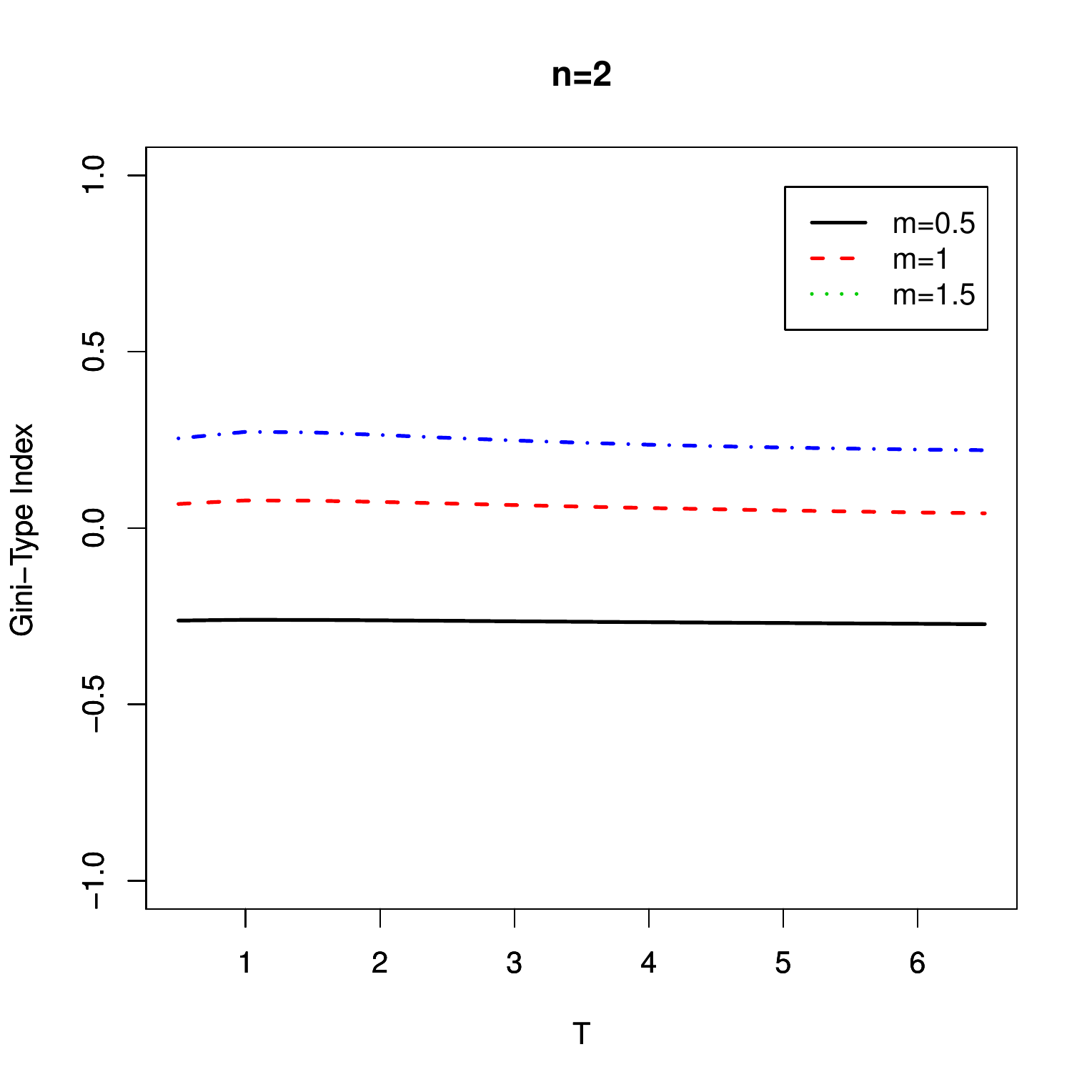}}
{\includegraphics[scale=0.3]{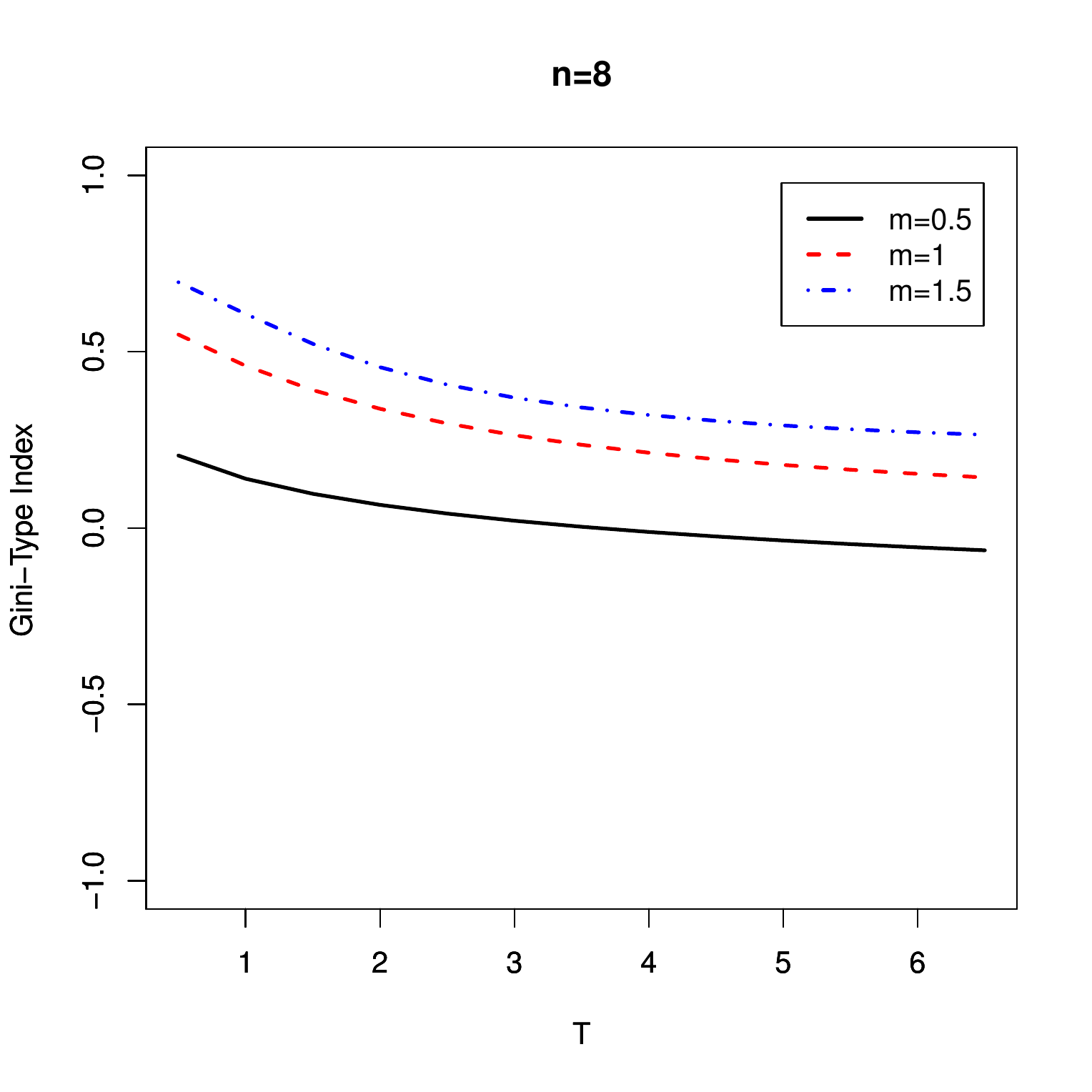}}\\
{\includegraphics[scale=0.3]{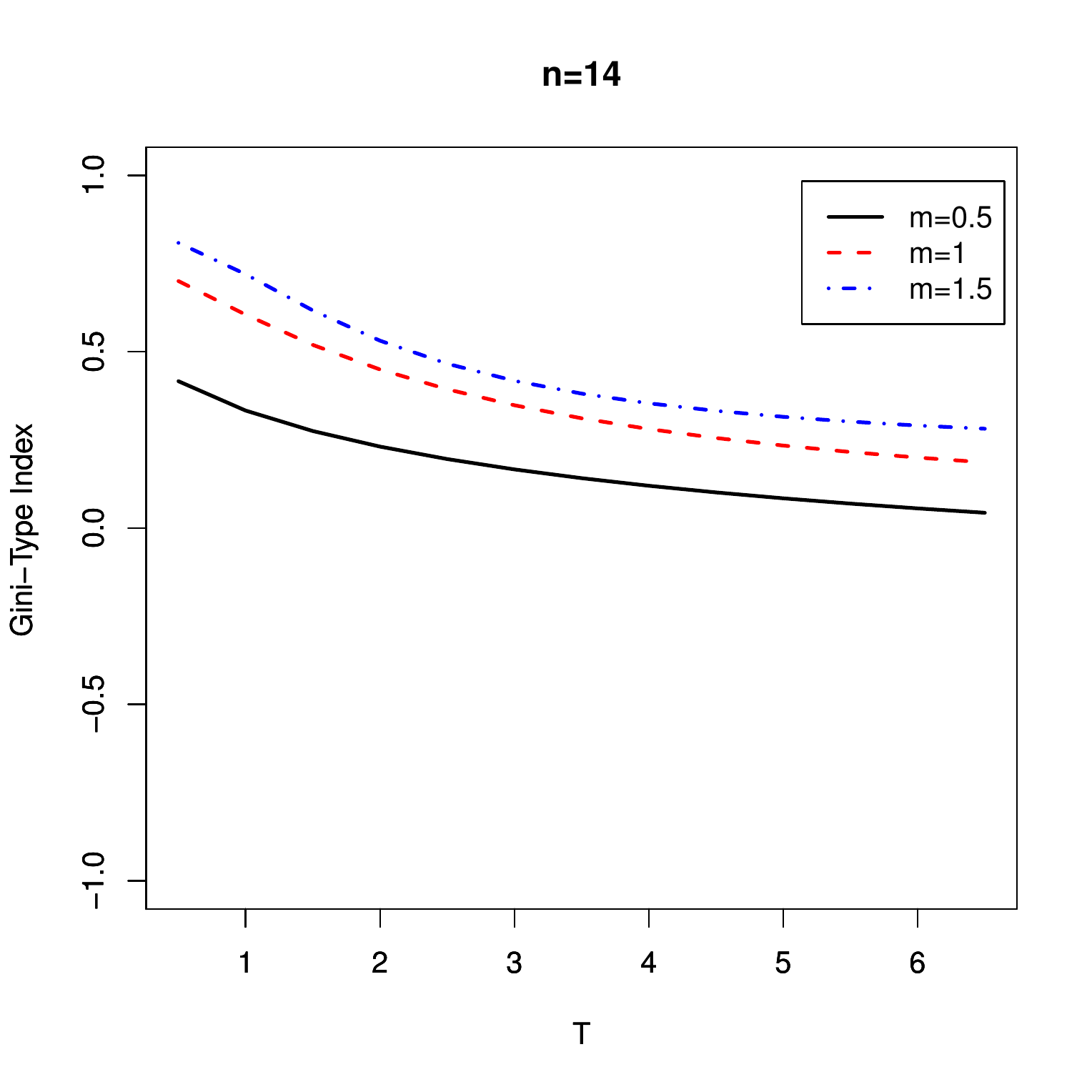}}
{\includegraphics[scale=0.3]{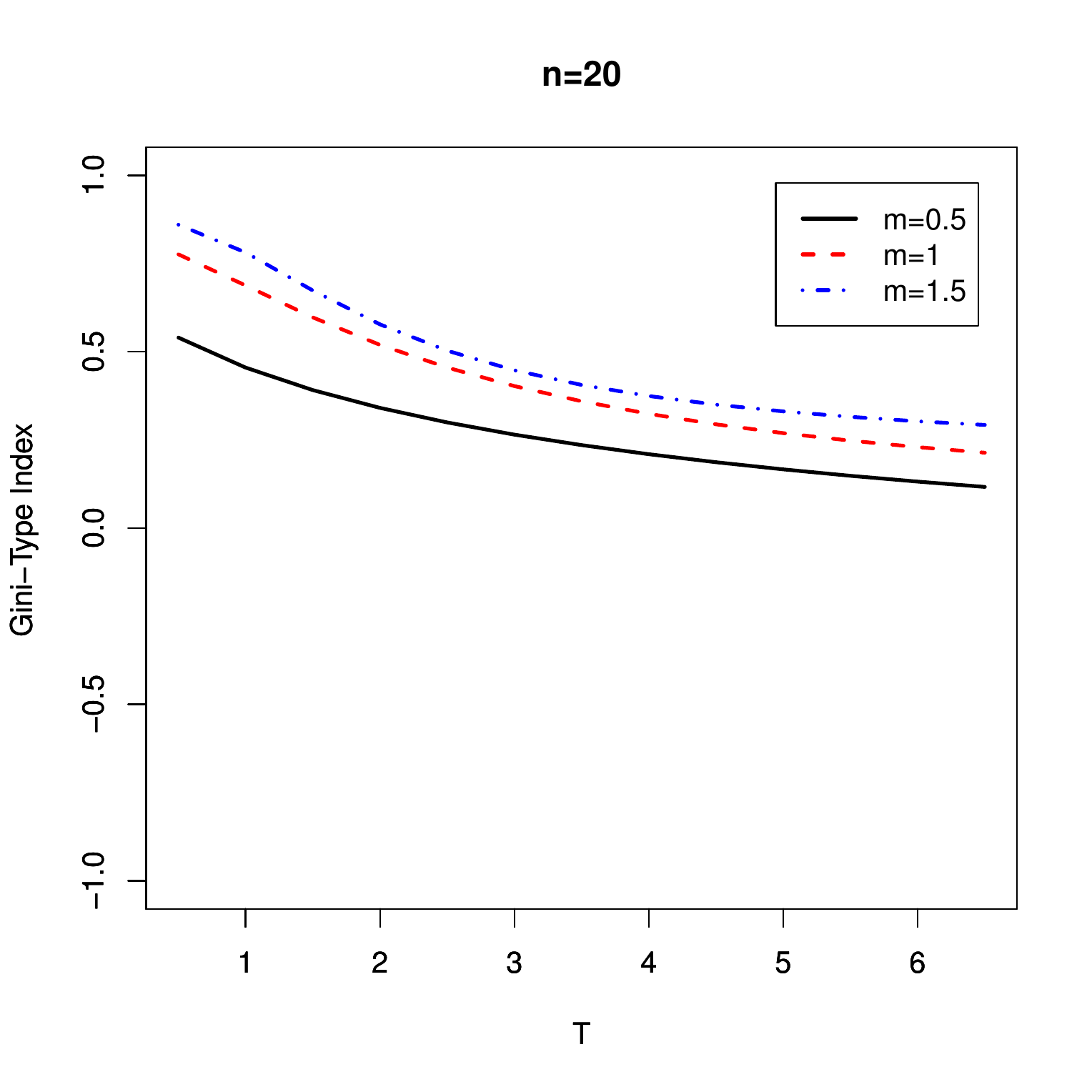}}
\caption{The GT index for the parallel-series system of Example \ref{Exm1} versus time while $\lambda=1$.}
\end{center}
\end{figure}\label{fig1}
\begin{figure}[ht]
\begin{center}
{\includegraphics[scale=0.3]{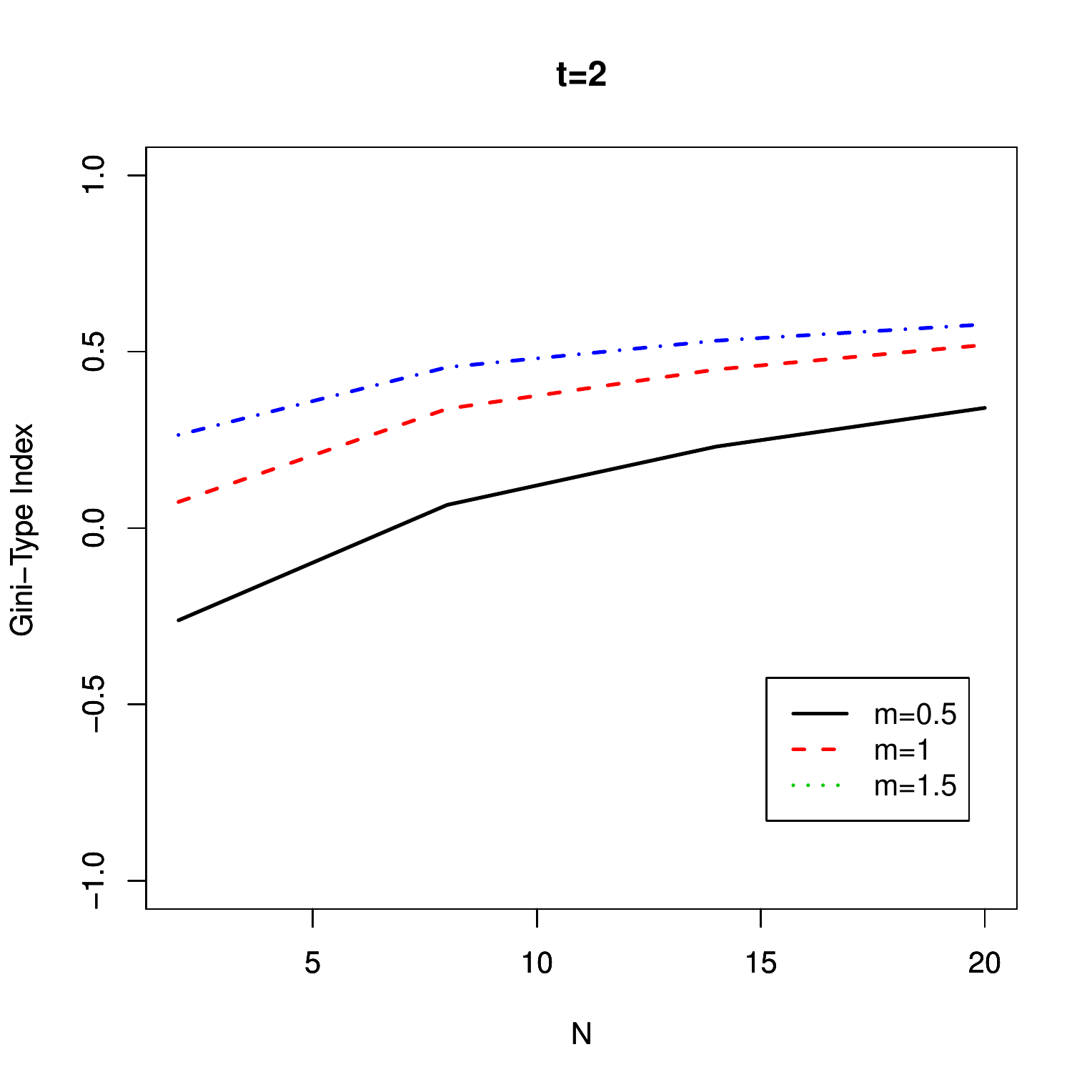}}
{\includegraphics[scale=0.3]{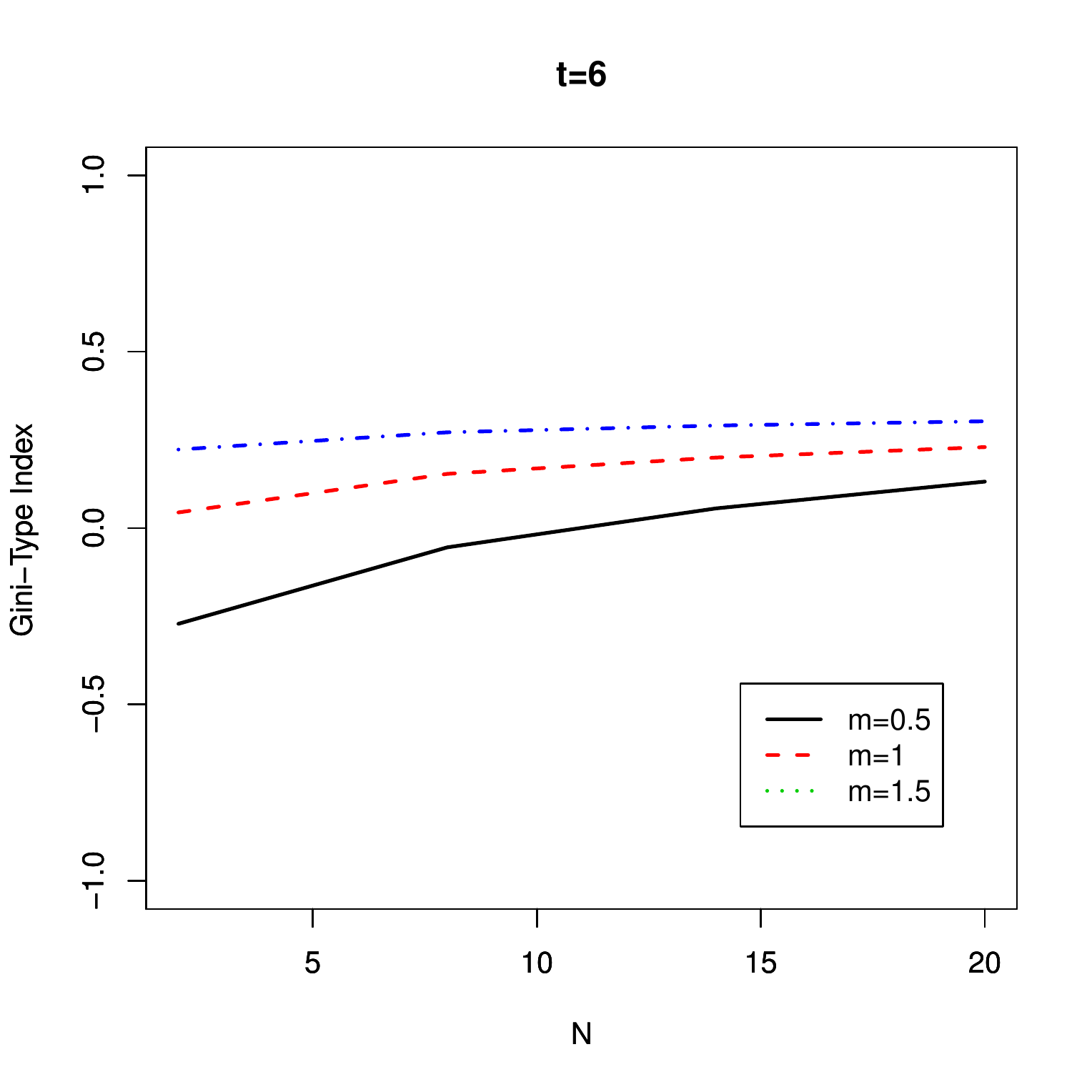}}
\caption{The GT index for parallel-series system of Example \ref{Exm1} versus the number $n$ 
of components while $\lambda=1$.}
\end{center}
\end{figure}\label{fig2}
\par
In general, the value of the GT index decreases in time and thus the intensity of system ageing is declined during the time; moreover, this behaviour is more severe as the number of components are increased. Figure 8 shows that the higher number of components 
increases the value of the GT index, which implies the more IFR property of the parallel-series system with shared components. 
\par
According to the Weibull distribution properties, when shape parameter satisfies $ 0<m<1 $, $ m=1 $ and $ m> 1$, the corresponding distribution is DFR, CFR and IFR, respectively. 
Considering Figure 8, the trend of GT indexes results that while the components are DFR and we have a limited number of components, the system will be DFR as well. But, CFR and IFR components always provide an IFR parallel-series system with shared components. 
\par
Figure 9 illustrates that the value of the GT index increases by increasing the number of components in the considered parallel-series system. Therefore, such a system is better to be constructed by less number of units to gain better ageing property.
\end{example}
\par
According to Example \ref{Exm1}, the ageing property of a complicated system can be 
determined by the GT index. This application is helpful in the planning of optimal systems and for 
efficient management. 
In the next section we discuss the dual of the system which has been studied in this section. 
\section{Series-parallel system with shared components}\label{sec6}
Under the assumptions of Section \ref{sec5}, here we define a series-parallel system with shared components 
having i.i.d. lifetimes with c.d.f $F(t)$. 
Let us suppose that each group of $ k $ components for fixed $ k $, $ 1 \leq k < n$, 
works as a local parallel system and we have 
\[ 
 \tilde Y_j=\max_{j \leq i \leq j+k-1} T_i, \qquad  j=1, 2,\dots, n-k+1, 
\]
which defines a local dependence among the components. 
The main system is constructed as the local parallel systems 
are connected in series and hence, the lifetime of the system is given by  
\[ 
T_{S-P}=\min_{1 \leq j \leq n-k+1}\tilde Y_j.
\]
As in the previous case, our objective is to obtain the c.d.f.\ of $T_{S-P}$, 
and then to analyse its ageing property by means of the GT index. 
Following the same procedure of Section \ref{sec5},  
the survival function of the series-parallel 
system with shared components is attained as
\begin{equation} \label{eq10}
{S}_{n-k+1}(t):=\mathbb P(T_{S-P}>t)
=\bar{F}(t) \sum^{k}_{i=1} F^{i-1}(t) {S}_{n-k+1-i}(t), \qquad t\in D_X^1.
\end{equation}
In general, determining the function ${S}_{n-k+1}(t)$ 
requires numerical methods to solve the difference equation of order $ k $ given in \eqref{eq10}. 
Then, the GT index of series-parallel system with shared components is given as follows 
\begin{equation}\label{eq11}
GT_{S-P}(t)=1-\frac{2 \int^t_0 \log(\bar{G}_{n-k+1}(u))du}{t \log(\bar{G}_{n-k+1}(t))}, 
\qquad t\in D_X^1.
\end{equation}
\subsection{Case $ k=2 $}
As a special case, we investigate the series-parallel system with shared components when 
$k=2$. See Figure 10 as a representation of such a system. 
According to \eqref{eq10}, for $n\geq 3$, $n\in \mathbb N$, 
the corresponding survival function can be written as
\begin{equation}\label{eqn12}
 {S}_{n-1}(t)=\bar{F}(t)  {S}_{n-2}(t)+\bar{F}(t) F(t)  {S}_{n-3}(t).
\end{equation}
To solve \eqref{eqn12}, we note that the auxiliary equation 
$$
 \beta ^{n-1}(t)-\bar{F}(t) \beta^{n-2}(t) -\bar{F}(t) F(t)\beta^{n-3}(t)=0,
$$
has the following solution:
\begin{equation}\label{eq14}
 {S}_n(t)=[1-D(t)]\beta^n_1(t)+D(t)\beta^n_2(t),\qquad t\in D_X^1,
\end{equation}
where 
$$ 
 \beta_{1,2}(t)=\frac{1}{2}\left[\bar{F}(t)\pm \sqrt{\bar{F}^2(t)+4\bar{F}(t)F(t)}\right], 
 \qquad  \beta_2(t) <0<\beta_1(t),
$$ 
with 
$$ 
 D(t)=\frac{\beta_{1}(t)+F^2(t)-1}{\beta_{1}(t)-\beta_{2}(t)}. 
$$
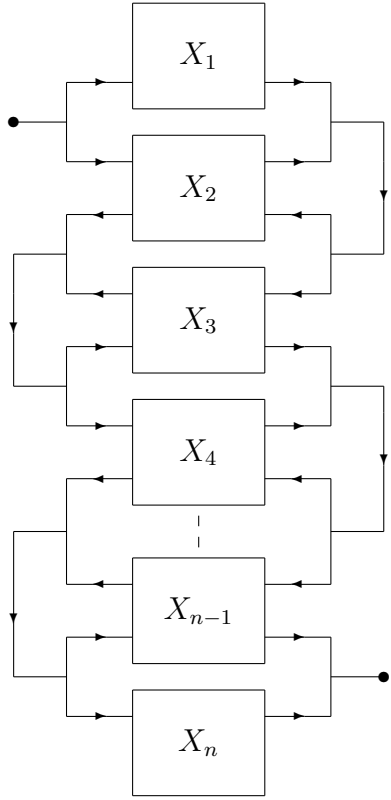
\begin{figure}[t]
\begin{center}
\begin{picture}(240,270)
\put(70,250){\makebox(20,15)[t]{\large $X_1$}}
\put(70,200){\makebox(20,15)[t]{\large $X_2$}}
\put(70,150){\makebox(20,15)[t]{\large $X_3$}}
\put(70,100){\makebox(20,15)[t]{\large $X_4$}}
\put(70,40){\makebox(20,15)[t]{\large $X_{n-1}$}}
\put(70,-10){\makebox(20,15)[t]{\large $X_{n}$}}
\put(55,240){\line(0,1){40}}
\put(55,190){\line(0,1){40}}
\put(55,140){\line(0,1){40}}
\put(55,90){\line(0,1){40}}
\put(55,30){\line(0,1){40}}
\put(55,-20){\line(0,1){40}}
\put(105,240){\line(0,1){40}}
\put(105,190){\line(0,1){40}}
\put(105,140){\line(0,1){40}}
\put(105,90){\line(0,1){40}}
\put(105,30){\line(0,1){40}}
\put(105,-20){\line(0,1){40}}
\put(55,280){\line(1,0){50}}
\put(55,240){\line(1,0){50}}
\put(55,230){\line(1,0){50}}
\put(55,190){\line(1,0){50}}
\put(55,180){\line(1,0){50}}
\put(55,140){\line(1,0){50}}
\put(55,130){\line(1,0){50}}
\put(55,90){\line(1,0){50}}
\put(55,70){\line(1,0){50}}
\put(55,30){\line(1,0){50}}
\put(55,20){\line(1,0){50}}
\put(55,-20){\line(1,0){50}}
\put(10,235){\circle*{4}}
\put(150,25){\circle*{4}}
\put(10,235){\line(1,0){20}}
\put(30,220){\line(0,1){30}}
\put(55,170){\vector(-1,0){15}}
\put(40,170){\line(-1,0){10}}
\put(55,200){\vector(-1,0){15}}
\put(40,200){\line(-1,0){10}}
\put(30,170){\line(0,1){30}}
\put(30,185){\line(-1,0){20}}
\put(10,185){\line(0,-1){15}}
\put(10,170){\vector(0,-1){15}}
\put(10,155){\line(0,-1){20}}
\put(30,150){\vector(1,0){15}}
\put(45,150){\line(1,0){10}}
\put(30,120){\vector(1,0){15}}
\put(45,120){\line(1,0){10}}
\put(105,150){\vector(1,0){15}}
\put(120,150){\line(1,0){10}}
\put(105,120){\vector(1,0){15}}
\put(120,120){\line(1,0){10}}
\put(130,150){\line(0,-1){30}}
\put(10,135){\line(1,0){20}}
\put(30,120){\line(0,10){30}}
\put(55,100){\vector(-1,0){15}}
\put(55,60){\vector(-1,0){15}}
\put(40,100){\line(-1,0){10}}
\put(40,60){\line(-1,0){10}}
\put(30,250){\vector(1,0){15}}
\put(30,220){\vector(1,0){15}}
\put(45,220){\line(1,0){10}}
\put(45,250){\line(1,0){10}}
\put(80,74){\line(0,1){4}}
\put(80,82){\line(0,1){4}}
\put(130,100){\vector(-1,0){15}}
\put(105,100){\line(1,0){10}}
\put(130,135){\line(1,0){20}}
\put(105,220){\vector(1,0){15}}
\put(105,250){\vector(1,0){15}}
\put(120,220){\line(1,0){10}}
\put(120,250){\line(1,0){10}}
\put(130,80){\line(1,0){20}}
\put(130,220){\line(0,1){30}}
\put(130,235){\line(1,0){20}}
\put(150,235){\line(0,-1){15}}
\put(150,220){\vector(0,-1){15}}
\put(150,205){\line(0,-1){20}}
\put(150,185){\line(-1,0){20}}
\put(130,200){\line(0,-1){30}}
\put(130,60){\vector(-1,0){15}}
\put(115,60){\line(-1,0){10}}
\put(130,200){\vector(-1,0){15}}
\put(115,200){\line(-1,0){10}}
\put(130,170){\vector(-1,0){15}}
\put(115,170){\line(-1,0){10}}
\put(150,125){\line(0,1){10}}
\put(150,125){\vector(0,-1){20}}
\put(150,80){\line(0,1){25}}
\put(130,100){\line(0,-1){40}} 
\put(30,60){\line(0,1){40}} 
\put(10,80){\line(1,0){20}} 
\put(45,40){\line(1,0){10}}
\put(30,40){\vector(1,0){15}}
\put(45,10){\line(1,0){10}}
\put(30,10){\vector(1,0){15}}
\put(30,10){\line(0,1){30}}
\put(120,40){\line(1,0){10}}
\put(105,40){\vector(1,0){15}}
\put(120,10){\line(1,0){10}}
\put(105,10){\vector(1,0){15}}
\put(130,10){\line(0,1){30}}
\put(10,25){\line(1,0){20}}
\put(130,25){\line(1,0){20}}
\put(10,80){\vector(0,-1){35}}
\put(10,25){\line(0,1){20}}
\end{picture}
\end{center}
\caption{Schematic representation of the series-parallel system with shared components 
when $ k=2 $ and $ n $ is even.}
\label{Fig:System2}
\end{figure}
\begin{example} \label{Exm2}
Suppose that the iid random lifetimes $  X_i $'s have Weibull c.d.f.\ as is mentioned in Example \ref{Exm1}. 
Thus, due to (\ref{eq14}), the survival function of the series-parallel system with shared components, for $k=2$, is given by 
\begin{equation}\label{eq15}
\begin{aligned}
 {S}_n(t)&=\frac{1}{B(t)}\left[1-(1-e^{-( \lambda t)^m} )^2-\frac{1}{2}e^{-(\lambda t)^m} +\frac{1}{2}B(t)\right]\\
&\times \frac{1}{2^n} \left[e^{-(\lambda t)^m}+B(t)\right]^n \\
&+\left\{1-\frac{1}{B(t)}\left[1-(1-e^{-(\lambda t)^m})^2-\frac{1}{2}e^{-(\lambda t)^m} +\frac{1}{2}B(t)\right]\right\} \\
&\times \frac{1}{2^n} \left[e^{-(\lambda t)^m}-B(t)\right]^n,
\qquad t>0,
\end{aligned}
\end{equation}
where
\[
B(t)=\sqrt{e^{-2(\lambda t)^m}+4 (1-e^{-(\lambda t)^m}) e^{-(\lambda t)^m}}.
\]
The values of the GT index for the considered system are illustrated in Figure 11, whereas, the GT index versus the number of components is presented in Figure 12.

\begin{figure}[ht] 
\begin{center}
{\includegraphics[scale=0.3]{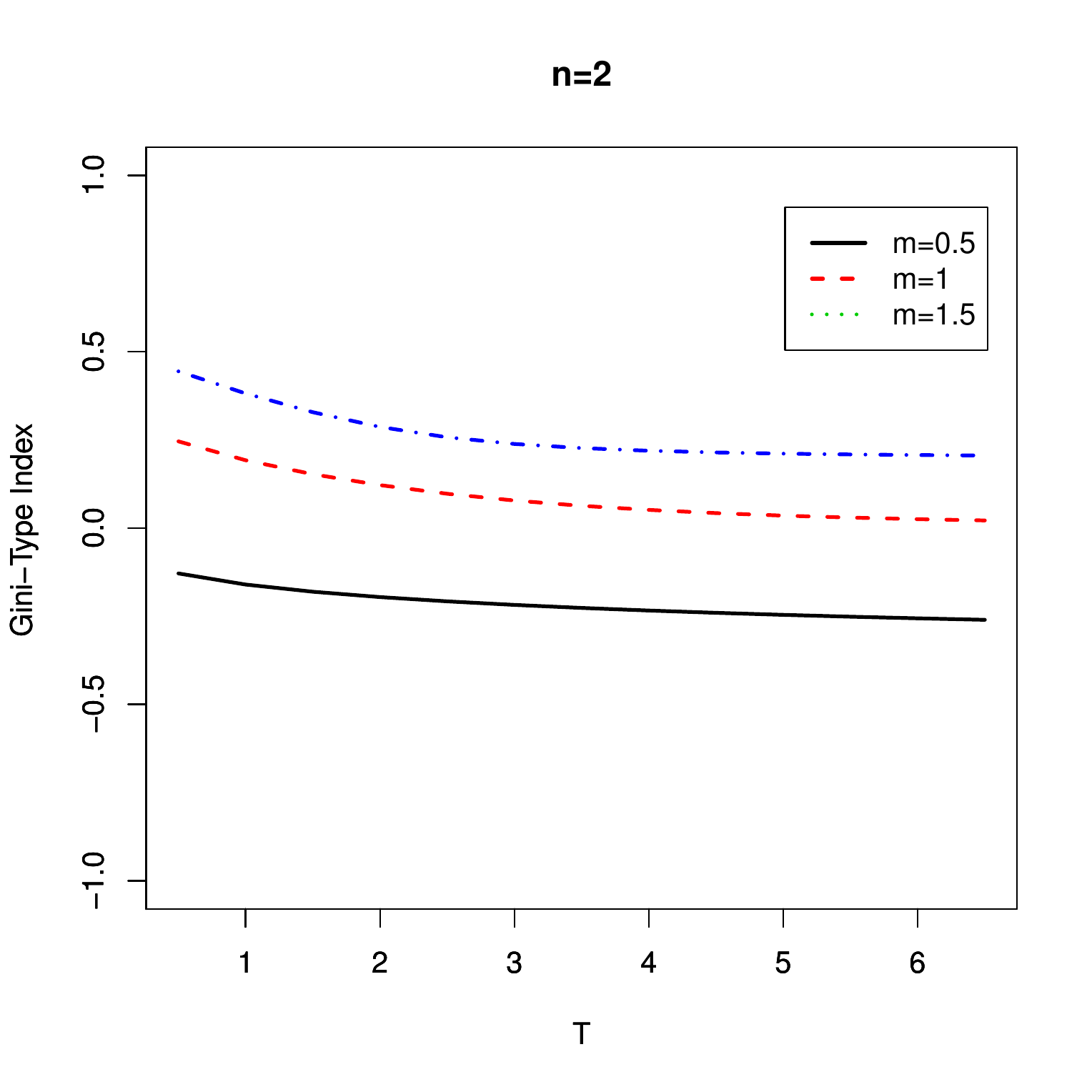}}
{\includegraphics[scale=0.3]{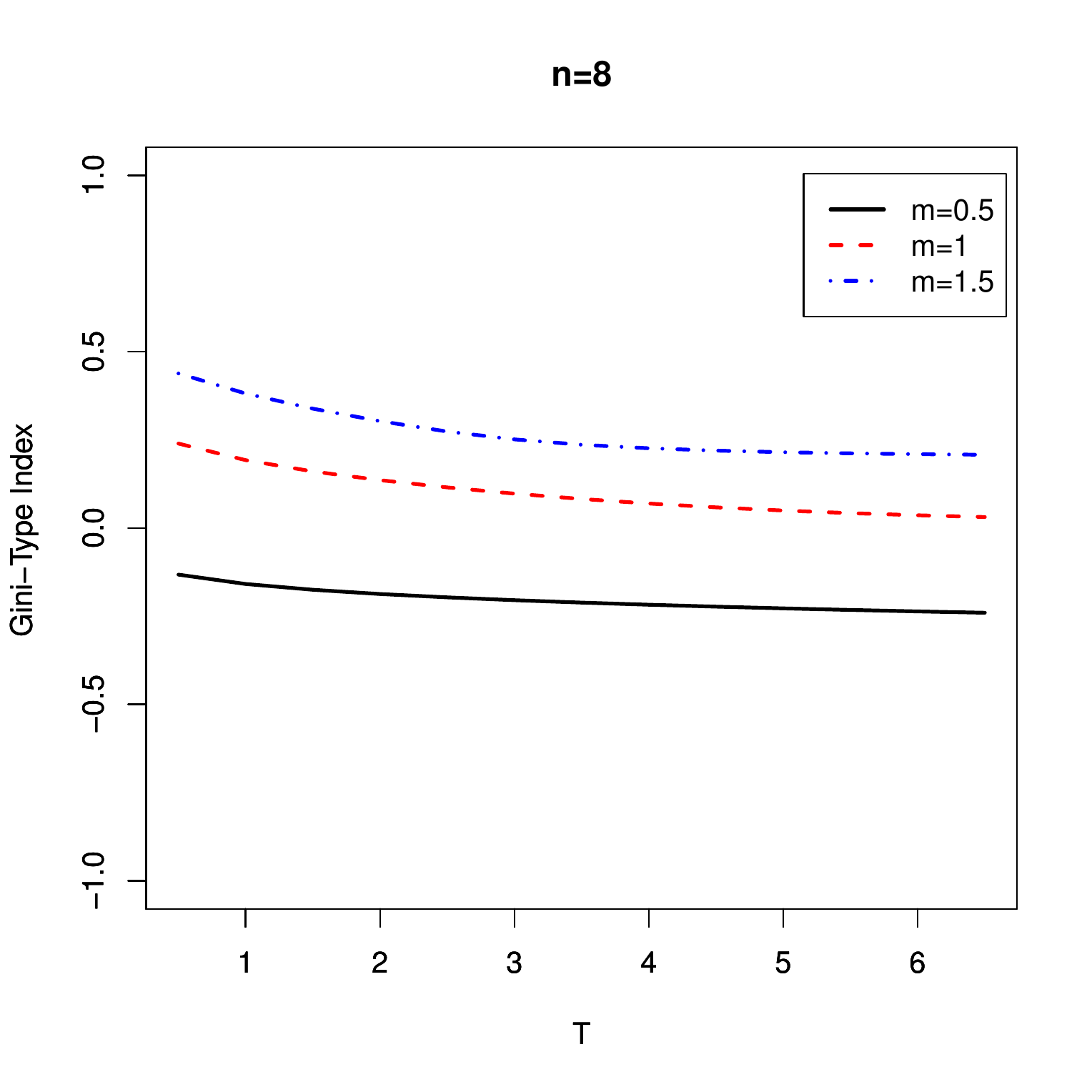}}\\
{\includegraphics[scale=0.3]{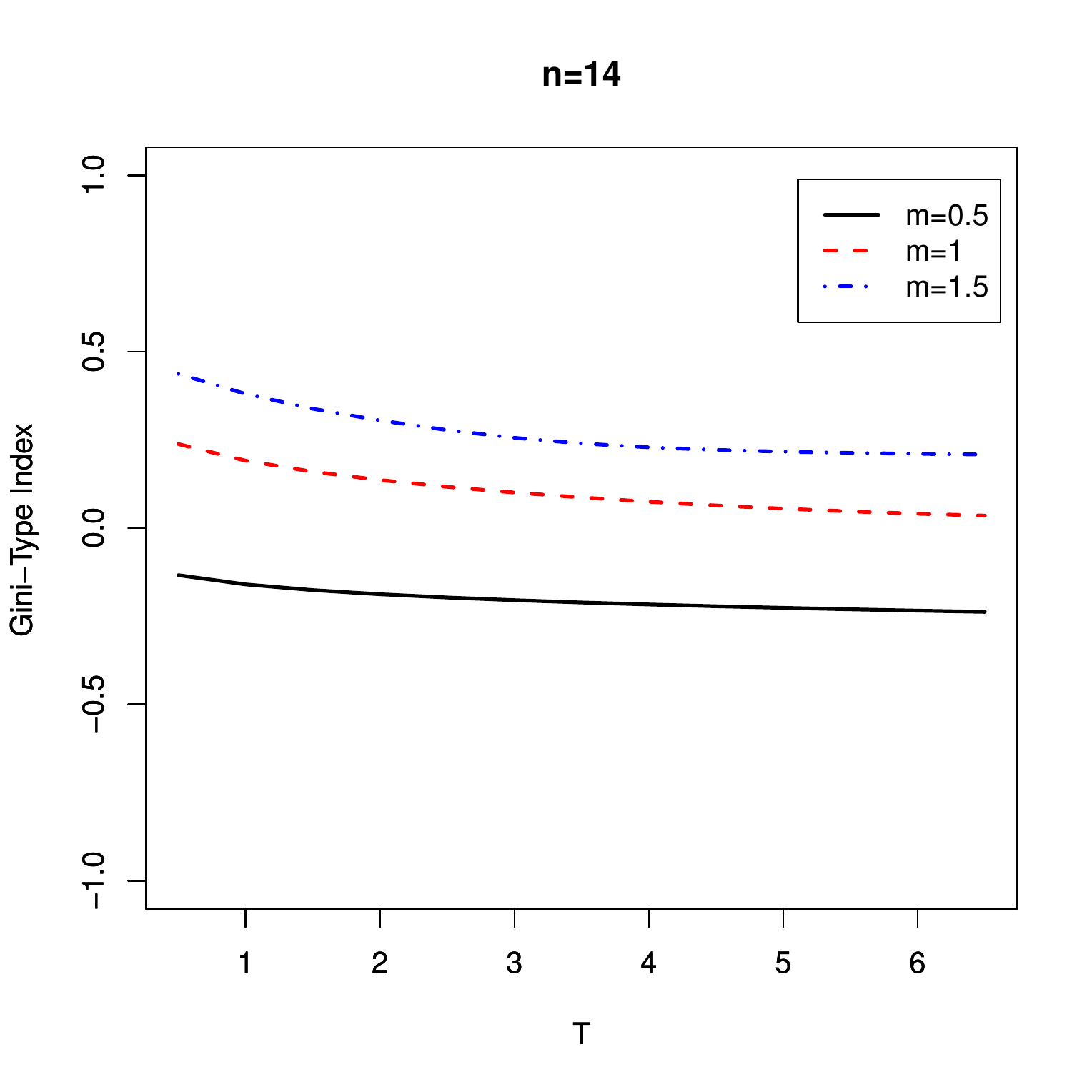}}
{\includegraphics[scale=0.3]{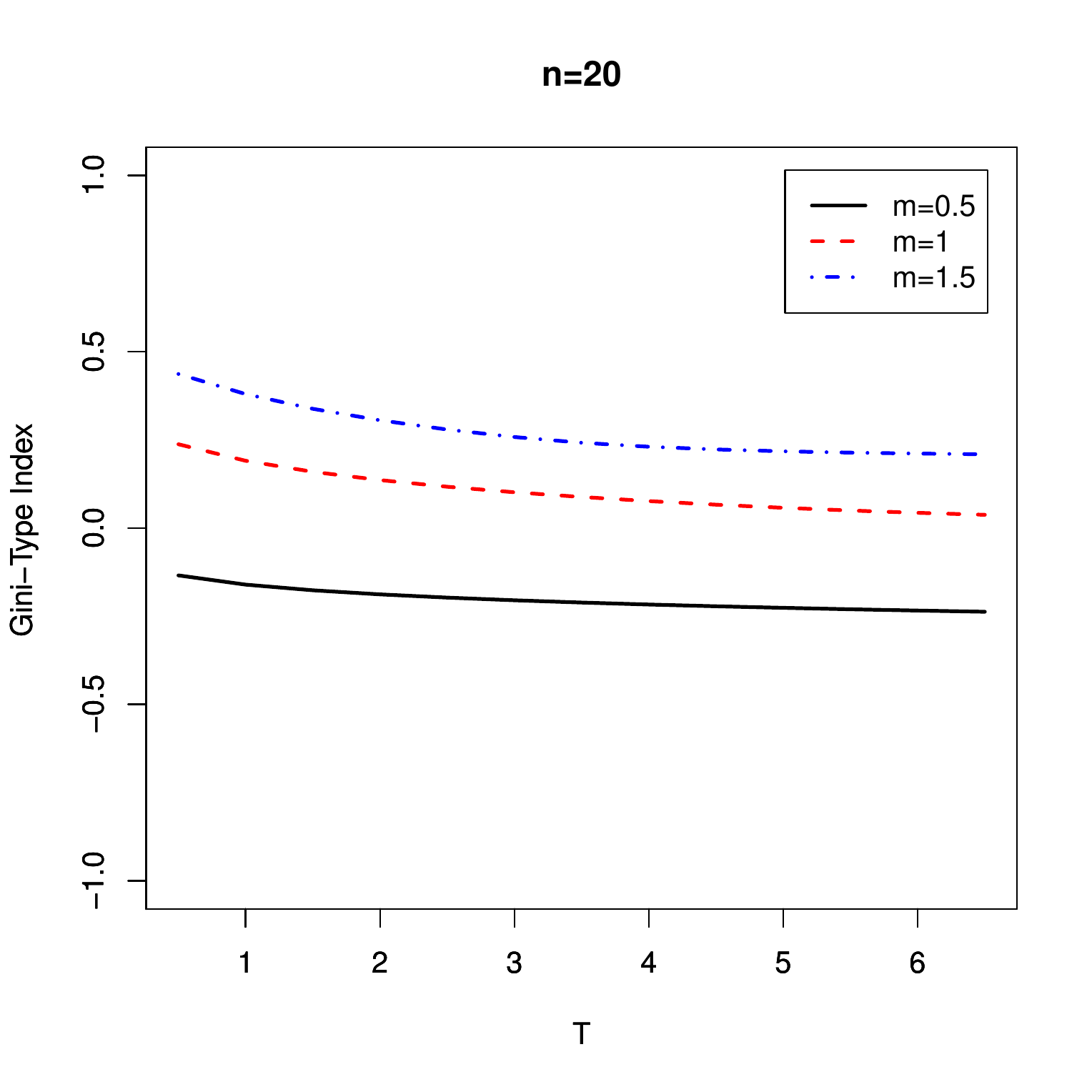}}
\caption{The GT index for the series-parallel system of Example \ref{Exm2} versus time, for $\lambda=1$.}
\end{center}
\end{figure} \label{fig3}
\begin{figure}[ht] 
\begin{center}
{\includegraphics[scale=0.3]{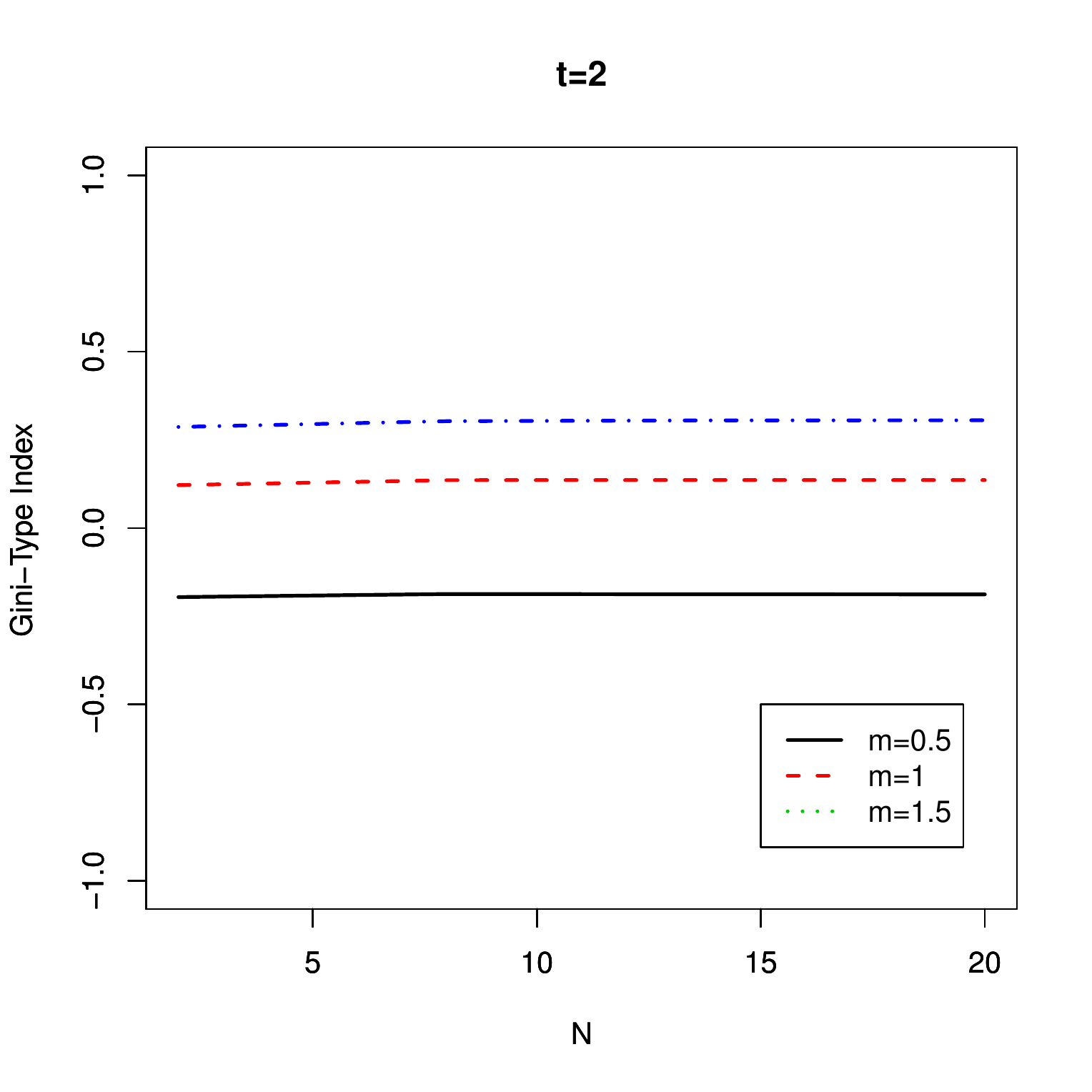}}
{\includegraphics[scale=0.3]{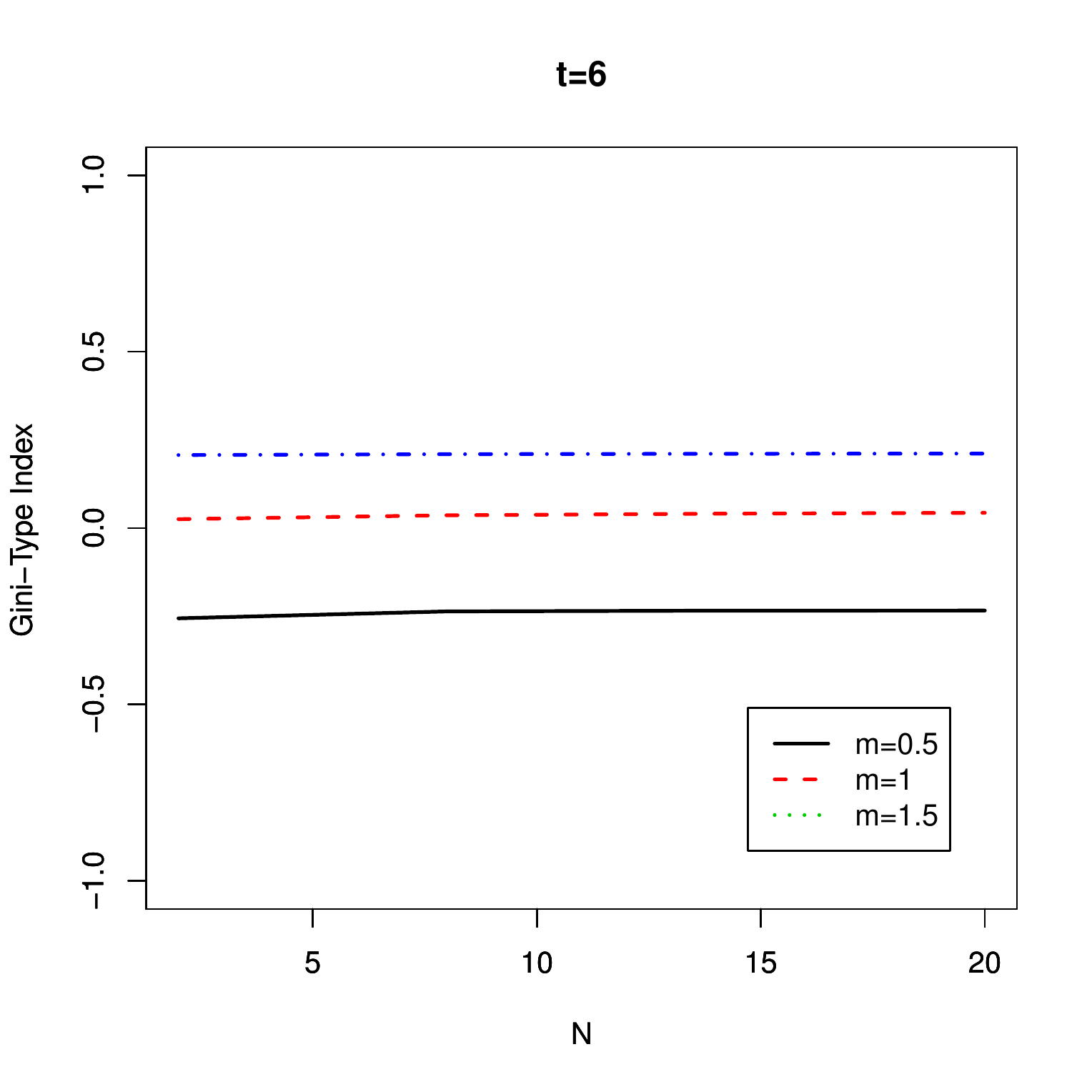}}
\caption{The GT index for the series-parallel system of Example \ref{Exm2} versus the number $n$ of components, for $\lambda=1$.}
\end{center}
\end{figure} \label{fig4}
\par
According to Figures 11 and 12, the GT index decreases in time and therefore it results less ageing property. It seems that increasing the number of components in the series-parallel system with  shared components has no significant influence on the GT index. Also, it is implied that when the components are DFR, the system is DFR as well. 
But, CFR and IFR components always provide an IFR series-parallel system with shared components. 
\end{example}
\par
We conclude this section by remarking that the GT index can be applied to compare the ageing properties of complex systems. In general, a parallel-series system with shared components takes larger values of GT index.  
So, considering the systems' ageing properties, parallel-series system with shared components deteriorates faster than the series-parallel system with shared components. 
Besides, as the number of components in the systems is increased,
the structure of parallel-series system reveals the more IFR property rather than the structure of series-parallel system. Although, when time increases the two systems' GT indexes slightly approach to 
the same fixed value and the graphs smoothly go flat. 
\section{Generalized Gini-type index}\label{sec7}

Shaked and Shanthikumar \cite{Shaked and Shanthikumar 1987}, \cite{Shaked and Shanthikumar 2014} introduced the multivariate conditional hazard functions and discussed the usefulness and some of their properties. Utilizing the definitions for multivariate cumulative hazard assists us to introduce the generalized GT (GGT) index. 
In this section, we shall see that the GGT index is particularly useful to compare ageing properties 
even of joint dependent random lifetimes. First, we give the definition for the bivariate case. The multivariate case is discussed later. 
\subsection{Bivariate Case}
Suppose that $ (X_1,X_2) \geq (0,0)$ is a random vector with a.c.\ joint survival function $ \bar{F}(t_1,t_2)=\mathbb P(X_1>t_1, X_2>t_2) $ and continuous joint d.f.\ $ f(t_1,t_2) $, representing the lifetime of two components. The following definition expresses the \textit{multivariate conditional hazard rate functions} in the bivariate case.
\begin{definition} \label{def6}
The failure rate of component $1 $, while both components are alive at time $ t$,
with $t \in \{s>0 : \bar{F}(s,s)>0\}$, is given by 
\begin{equation} \label{eq17}
\begin{aligned}
h_{1}(t)&=\lim_{\Delta t \downarrow 0 } 
\dfrac{\mathbb P(t<X_1\leq t +  \Delta t ~|~ X_1>t, X_2>t)}{ \Delta t}\\
&=\dfrac{-\frac{\partial}{\partial t_1}\bar{F}(t_1,t) |_{t_1=t}}{\bar{F}(t,t)}.
\end{aligned}
\end{equation}
On the other hand, the failure rate of component $ 1 $, given that it is alive at time $ t $ and component 
$2$ has already failed sometime earlier at time $ t_2$, is
\begin{equation} \label{eq18}
\begin{aligned}
h_{1|2}(t \,|\, t_2)&=\lim_{ \Delta t \downarrow 0 } \dfrac{\mathbb P(t<X_1 \leq t+ \Delta t ~|~ X_1 >t, X_2=t_2)}{ \Delta t}\\
&=\dfrac{f(t,t_2)}{-\frac{\partial}{\partial t_2}\bar{F}(t,t_2) },
\end{aligned}
\end{equation}
for $t \in \{s \geq t_2: \frac{\partial}{\partial t_2}\bar{F}(s,t_2) <0 \}$, and $t_2 \in \{s>0:\bar{F}(s,s) >0\}$. 
\end{definition}
\par
The same definition of failure rate for component $ 2 $ can be given similarly 
(for more details see \cite{Shaked and Shanthikumar 1987}, 
\cite{Shaked and Shanthikumar 2014}).
\par
The total accumulated hazards for the aforementioned components can be reached as well.
\begin{definition}\label{def7}
The hazard accumulated by component $ i $ by time $ t >0 $, given that it was alive during the time interval 
$(0,t]$, is attained as
\begin{equation} \label{eq20}
H_{i}(t)=\int^t_0 h_{i}(u)du, \qquad i=1, 2.
\end{equation}
The cumulative hazard of component $1$, given that it is alive after the failure of 
component $2$, is achieved as \begin{equation}\label{eq21}
\begin{aligned}
H_{1 | 2}(t-t_2 \,|\, t_2)&=\int^t_{t_2} h_{1 | 2}(u \,|\, t_2) du \\
&=-\log \left(-\frac{\partial}{\partial t_2}\bar{F}(u,t_2)\right) \Big|_{u=t-t_2}, \qquad  0<t_2<t.
\end{aligned}
\end{equation}
The cumulative hazard of component $2$, given that it is alive after the failure of 
component $1$,  can be defined similarly. 
\end{definition}
One needs to note that $ H_{1}(t_1) $ is the hazard rate accumulated by component $ 1$ up to time $ t_1 $, 
given $t_1 \leq t_2 $. Similarly, $ H_{1}(t_2) $ is the hazard accumulated by component $ 1$ by time  $t_2$, 
given that $ t_1 > t_2 $. Clearly, $ H_{1 | 2}(t_1-t_2\,|\,t_2) $ is the hazard accumulated by component $ 1 $ over time interval $ (t_2,t_1]$ while $ t_1 > t_2 $.
Therefore, the total hazard accumulated by component $ 1 $ by the time it is failed, is
\begin{align} \label{eq22}
\Lambda_1 (t_1,t_2)=
\left\{
\begin{array}{ll}
 H_{1}(t_1),  & \hbox{if }t_1 \leq t_2  \\
 H_{1}(t_2)+H_{1 | 2}(t_1-t_2 \,|\, t_2), & \hbox{if }t_1> t_2,
\end{array}
\right.
\end{align}
which depends on the actual value of $ t_2 $. The total hazard accumulated by component $2$ before its failure can be attained similarly.
\par
According to \eqref{eq22}, the GGT index of a given component,  
depending on the lifetime of the other, is defined as follows, where we assume that 
$$ 
 D^1_{X_1 | t_2}=\left\{t>t_2:  0<-\frac{\partial}{\partial t_2}\bar{F}(t,t_2)<1\right\}.
$$
Since $\Lambda_1 (t_1,t_2)$ has different expressions for $t_1 \leq t_2$ and 
$t_1> t_2$, also for the GGT index we have two cases. 
\begin{definition}\label{def8}
Suppose that the non-negative a.c.\ random vector $(X_1,X_2)$ represents the 
lifetime of two components. The GGT index of component $1$ in $(0,t_1]$,  
while $ t_2 $ is the lifetime of components $ 2 $, is given hereafter: \\
$\bullet$ \ if $ 0<t_1 \leq t_2$, with $t_1\in D^1_{X_1}$, 
$$
 GGT_{1}(t_1,t_2)=1-\dfrac{2\int^{t_1}_0  H_{1}(u) du}{ t_1  H_{1}(t_1)};
$$
$\bullet$ \ if $ t_1 > t_2$, with $t_2\in D^1$ and $t_1 \in D^1_{X_1 | t_2}$, 
$$
 GGT_{1}(t_1,t_2)
 =1-\dfrac{2   \left[ \int^{t_2}_0 H_{1}(u)du +(t_1-t_2)H_{1}(t_2) +\int^{t_1}_{t_2} H_{1 | 2}(u-t_2 \,|\, t_2) du\right]}{t_1 \left[ H_{1}(t_2)+ H_{1 | 2}(t_1-t_2 \,|\, t_2)\right]}.
$$
\end{definition}
\par
The GGT index for component 2 in $ (0,t_2] $, depending on the lifetime of 
components 1, can be defined similarly.
\par
In case that one of the components has failed earlier, we define a conditioned GGT index which considers the live component behaviour just after the failure of the other component.

\begin{definition}\label{def9}
Assume that non-negative a.c.\ random vector $ (X_1,X_2) $ is the lifetime of two components. 
The conditioned GGT index for component 1 after the failure of component 2 in $ (t_2,t_1]$, for 
$t_1 \in  D^1_{X_1 | t_2}$   is derived by
\begin{equation}
GGT_{1 | 2}(t_1\,|\,t_2)=1-\dfrac{2 \int^{t_1}_{t_2} H_{1 | 2}(u-t_2 \,|\, t_2) du}{(t_1-t_2)H_{1 | 2}(t_1-t_2 \,|\, t_2)}. 
\end{equation}
\end{definition}
\par
The conditioned GGT index for component 2 in $ (t_1,t_2] $ can be defined similarly.
For a better intuition about the aforementioned GGT indexes see Figure 13, where 
the plot of $\Lambda_1(t_1,t_2)$ is given for the joint bivariate Pareto distribution studied in case (II) of Example \ref{exmp4}, when $t_1\in [0,1]$ and $t_2=0.35$. For a geometric interpretation, 
note that the GGT index introduced in Definition \ref{def7} can be viewed 
as the ratio of the indicated areas, $B/(A+B)$ for $0<t_1\leq t_2$, and $(B+D)/(A+B+C+D)$ for $t_1>t_2$. 
Figure 13 should be compared to Figure 10.1 of \cite{Kaminskiy and Krivtsoz 2010} in order to establish the difference between the univariate case and the present bivariate case. 
\begin{figure}
\begin{center}
\includegraphics[scale=.9]{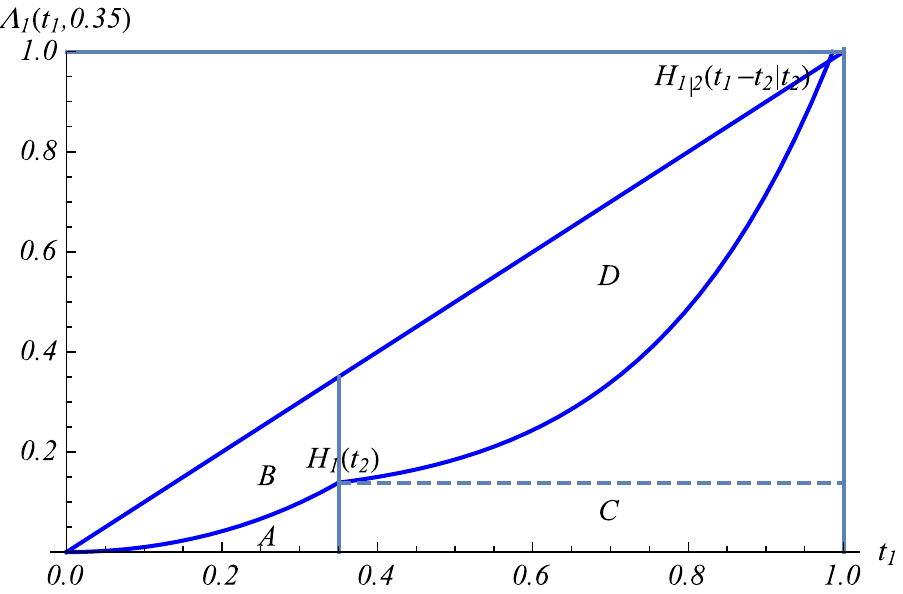}
\end{center}
\caption{Plot of $\Lambda_1(t_1,t_2)$ for the case (II) of Example \ref{exmp4}, and regions defining the 
GGT index.}
\end{figure}
%
\begin{remark}\label{rem2}
According to Definitions \ref{def1}, \ref{def8} and \ref{def9}, the following relations are concluded:
\begin{itemize}
\item For $ t_1 \leq t_2 $ we have 
\begin{equation}
 GGT_{1}(t_1,t_2)=GT_{X_1}(t_1),
\end{equation}
and therefore all the properties for $ GT_{X_1}(t_1) $ hold for $  GGT_{1}(t_1,t_2) $ as well.

\item For $ t_1 > t_2 $ it is derived that
\begin{equation}
\begin{aligned}
GGT_{1}(t_1,t_2)&=1- \frac{1}{t_1 \left[ H_{1}(t_2)+ H_{1 | 2}(t_1-t_2 \,|\, t_2)\right]} \lbrace t_2 H_1(t_2) (1-GT_{X_1}(t_2)) \\
&+(t_1-t_2)\left[ 2h_1(t_2)+H_{1|2}(t_1-t_2 \ | \ t_2)(1-GGT_{1|2}(t_1-t_2 \ | \ t_2)) \right]  \rbrace,
\end{aligned}
\end{equation}
by which $ GGT_{1}(t_1,t_2) $ is a linear combination of $ GT_{X_1}(t_2) $ and $ GGT_{1 | 2}(t_1 \ | \ t_2) $.
\end{itemize}
\end{remark}
\par

Obviously, as $ t_1 < t_2 $ one has the following properties 
of the generalized GT index, based on the monotonicity of the 
conditional hazard rate $h_{1|2}$: 
\\ 
$\bullet$ \  $GGT_{1 | 2}(t_1 \,| \,t_2) >0 $ when  $h_{1|2}(t_1 \,|\, t_2)$ is increasing; \\
$\bullet$ \ $GGT_{1 | 2}(t_1 \,| \,t_2) < 0 $ when  $h_{1|2}(t_1 \,|\, t_2)$ is decreasing; \\
$\bullet$ \ $GGT_{1 | 2}(t_1 \,| \,t_2) = 0 $ when  $h_{1|2}(t_1 \,|\, t_2)$ is constant.


In the rest of this section, some examples of GGT indexes are provided for better intuition of the aforementioned definitions. 
\begin{example}\label{exmp4}
Let us consider various choices of the joint survival function of  $(X_1,X_2)$. 
 \begin{itemize}
\item[(I)]
For the joint Bivariate Pareto survival function 
\[
\bar{F}(t_1,t_2)=(1+t_1+t_2)^{-1}, \qquad t_1, t_2 > 0,
\]
it is easily reached that
\begin{align*}
H_{1}(t)&=\frac{1}{2} \log (1+2 t), \qquad t \geq 0, \cr
H_{1 | 2}(t \,|\, t_2)&=2 \log\left(1+\frac{t}{1+2 t_2}\right), \qquad t,\ t_2> 0, \cr
\end{align*}
According to Definition \ref{def8}, we have 
$$
GGT_1(t_1,t_2)=-1-\frac{1}{t_1}+\frac{2}{\log(1+2 t_1)}, \quad t_1 \leq t_2, \ t_1 \in D^1_{X_1},
$$
and 
$$
GGT_1(t_1,t_2)=\dfrac{8(t_1-t_2)-4(2+t_1+3t_2) \log\left(\frac{1+t_1+t_2}{1+2t_2}\right)+t_2 \log\left(1+2t_2\right)}{4(t_1-t_2) \log\left(\frac{1+t_1+t_2}{1+2t_2}\right)+t_2\log\left(1+2t_2\right)}, 
$$
for $t_2 \in D^1_{X_1}, \ t_1 \in D^1_{X_1 | t_2}$. 
The conditional GGT index will be given as follows
\begin{align}
GGT_{1|2}(t_1\,|\,t_2)&=1+8 \log \left(\frac{1+t_1+t_2}{1+2t_2}\right) \nonumber \cr
&-8\frac{1+t_1+t_2}{t_1-t_2} \log\left(\frac{1+t_1+t_2}{1+2t_2}\right)^2, \qquad t_1 > t_2, \ t_1 \in D^1_{X_1 | t_2}.\nonumber
\end{align}

\item[(II)]
For the joint survival function 
 \[
\bar{F}(t_1,t_2)=\exp\{1-\exp(t^2_1+t^2_2)\}, \qquad t_1, t_2 >0, 
\]
we have 
\begin{align*}
H_{1}(t)&=\frac{1}{2} (-1+e^{2t^2}), \qquad t \geq 0, \cr
H_{1 | 2}(t \,|\, t_2)&=-e^{2t^2}+e^{t^2+t^2_2}-t^2+t^2_2, \qquad t,\ t_2> 0.
\end{align*}
Thus, the GGT index is 
$$
GGT_1(t_1,t_2)=\coth(t^2_1)-\frac{1}{2t_1}\sqrt{\frac{\pi}{2}}(-1+\coth(t^2_1)){\rm Erfi}(\sqrt{2}t_1), \quad t_1 \leq t_2, \ t_1 \in D^1_{X_1},
$$
where $ {\rm Erfi}(z) $ is imaginary error function $\frac{{\rm erfi}(iz)}{i}$. Also, 
for $t_2 \in D^1_{X_1}$ and $t_1 \in D^1_{X_1 | t_2}$ we have
\begin{align}
GGT_1(t_1,t_2)&=1+\bigg[4\left(-3t_1-2t^3_1 +6 t_1 t^2_2-4t^3_2+3e^{2 t^2_2} (-t_1+t_2)\right) \cr
&+12e^{2t^2_2}\sqrt{\pi}({\rm Erfi}(t_1)-{\rm Erfi}(t_2))+3\sqrt{2\pi} {\rm Erfi}(\sqrt{2}t_2)\bigg]\cr
&  \times \big[6t_1(1+e^{2t^2_2}-2e^{t^2_1+t^2_2}+2t^2_1-2t^2_2)\big]^{-1} .
\nonumber
\end{align}
The conditioned GGT index is derived by
\begin{align}
GGT_{1 | 2}(t_1 \, | \, t_2)&=1+\frac{1}{3(t_1-t_2)} 2 (e^{2t^2_2}+e^{t^2_1+t^2_2}-t^2_1+t^2_2) \cr
& \times \big[ (t_1-t_2)^2 (t_1+2 t_2) - 3 e^{t^2_1+t^2_2} DF(t_1) \cr
&+3 e^{2t^2_2} (t_1-t_2) + DF(t_2)\big],  \qquad t_1 > t_2, \ t_1 \in D^1_{X_1 | t_2},
\nonumber
\end{align}
where $ DF(x)=e^{-x^2} \int^x_0 e^{y^2}dy=\sqrt{\pi}e^{-x^2} {\rm Erfi}(x)/2$ is the Dawson function.

\item[(III)]
For the joint survival function 
 \[
\bar{F}(t_1,t_2)=(1+t^3_1+t^3_2)^{-2}, \qquad  t_1, t_2 >0,
\]
we have 
\begin{align*}
H_{1}(t)&=\log\left(1+2t^3\right), \qquad t \geq 0, \cr
H_{1 | 2}(t \,|\, t_2)&=3 \log\left(\frac{1+t^3+t^3_2}{1+2t^2_2}\right), \qquad t,\ t_2> 0. \cr
\end{align*}
The GGT indexes for component 1 are derived as the follows:
$$
GGT_1(t_1,t_2)=-1+\dfrac{6-6 \ {}_2F_1 (1,\frac{1}{3};\frac{3}{4};-2t^3_1)}{\log\left(1+2t^3_1\right)}, \quad t_1 \leq t_2, \ t_1 \in D^1_{X_1},
$$
where ${}_2F_1(a,b;c;z) $ is the Hypergeometric function. Moreover, one has 
\begin{align}
GGT_1(t_1,t_2)&=5- \bigg\lbrace 6 \bigg[ -3t_1+2t_2+t_2 \ {}_2F_1\left(1,\frac{1}{3};\frac{4}{3};-2t^3_2\right) \cr
&-3 \ t_2 \ {}_2F_1\left(1,\frac{1}{3};\frac{4}{3};\frac{-t^3_2}{1+t^3_2}\right) +t_1  
\bigg(3 \ {}_2F_1\left(1,\frac{1}{3};\frac{4}{3};\frac{-t^3_1}{1+t^3_2}\right) \cr
&+\log\left(1+t^3_2\right)+3\log\left(\frac{1+t^3_1+t^3_2}{1+2t^3_2}\right) \bigg) \bigg]  \bigg\rbrace, \quad t_2 \in D^1_{X_1}, \  t_1 \in D^1_{X_1 | t_2}.
\nonumber
  \end{align}
The conditioned GGT index is given as 
 \begin{align}
 GGT_{1|2}(t_1\, | \,t_2)&=1+54 \log\left(\frac{1+t^3_1+t^3_2}{1+2t^3_2}\right)+\frac{1}{t_1-t_2}+ \bigg\lbrace 9\log\left(\frac{1+t^3_1+t^3_2}{1+2t^3_2}\right)  \cr 
& \times \bigg[-2t_1\log\left(\frac{1+t^3_1+t^3_2}{1+2t^3_2}\right) \cr &+(1+t^3_2)^{\frac{1}{3}} \bigg( 2\sqrt{3}\arctan\left(\frac{1}{\sqrt{3}} (1-\frac{2t_1}{(1+t^3_2)^{\frac{1}{3}}})\right) \cr
 &-2\sqrt{3}\arctan\left(\frac{1}{\sqrt{3}} (-1+\frac{2t_2}{(1+t^3_2)^{\frac{1}{3}}})\right) \cr
&-2 \log\left(t_1+(1+t^3_2)^{\frac{1}{3}}\right)+2 \log\left(t_2+(1+t^3_2)^{\frac{1}{3}}\right) \cr
&+\log\left(t^2_1- t_1(1+t^3_2)^{\frac{1}{3}}+(1+t^3_2)^{\frac{2}{3}}\right) \cr
&-\log\left(t^2_2- t_2(1+t^3_2)^{\frac{1}{3}}+(1+t^3_2)^{\frac{2}{3}}\right) \bigg) \bigg] \bigg\rbrace, \quad t_1 \in D^1_{X_1 | t_2}. 
\nonumber
 \end{align}
\end{itemize}

Figures 14--16 illustrate the trends of GGT indexes versus components lifetimes for the 
three cases treated in Example \ref{exmp4}.

\begin{figure}[ht] 
\begin{center}
{\includegraphics[scale=0.55]{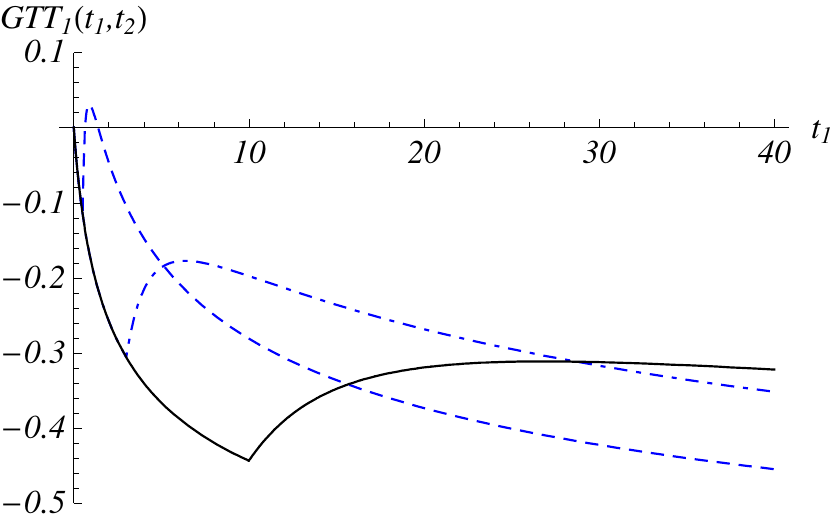}}
{\includegraphics[scale=0.55]{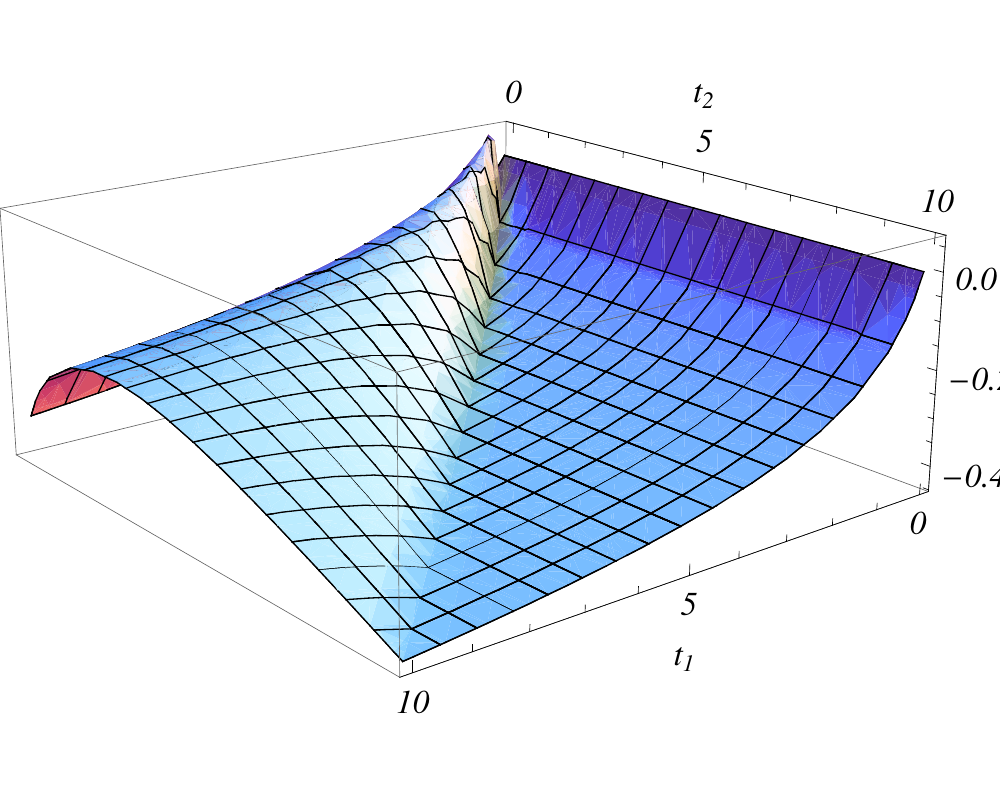}}
\caption{GGT index of component $ 1 $ when $ t_1 \leq t_2 $ and $ t_2= 0.5, 3, 10$ (left) and  $ t_1 > t_2 $ (right), for the joint Bivariate distribution given in Example \ref{exmp4} (I).}
\end{center}
\end{figure} \label{fig12}

According to Figure 14 for the case (I), the GGT index of component $ 1 $ is negative for large values of $ t_1 $. When both components are alive, the GGT index is decreasing as $ t_1 $ is increased. If it is supposed that $ t_2 $ is failed earlier, 
the GGT index decreases according to $ t_1 $ and $ t_2 $.

\begin{figure}[ht] 
\begin{center}
{\includegraphics[scale=0.55]{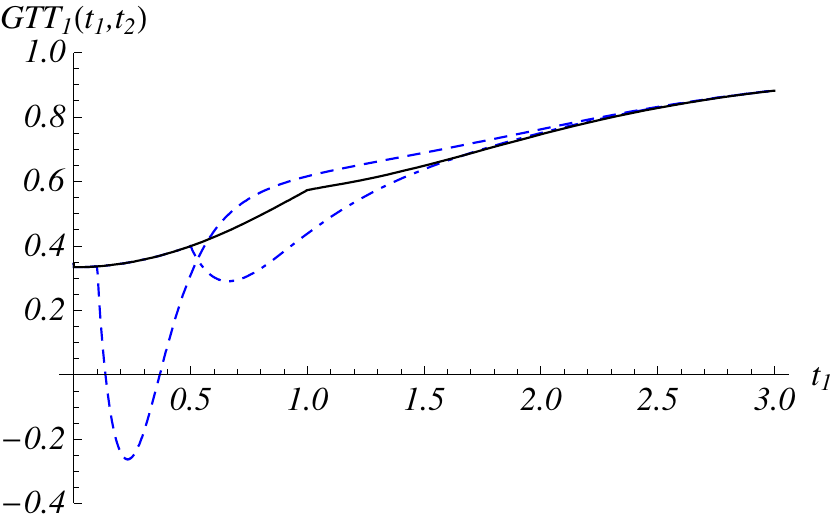}}
{\includegraphics[scale=0.55]{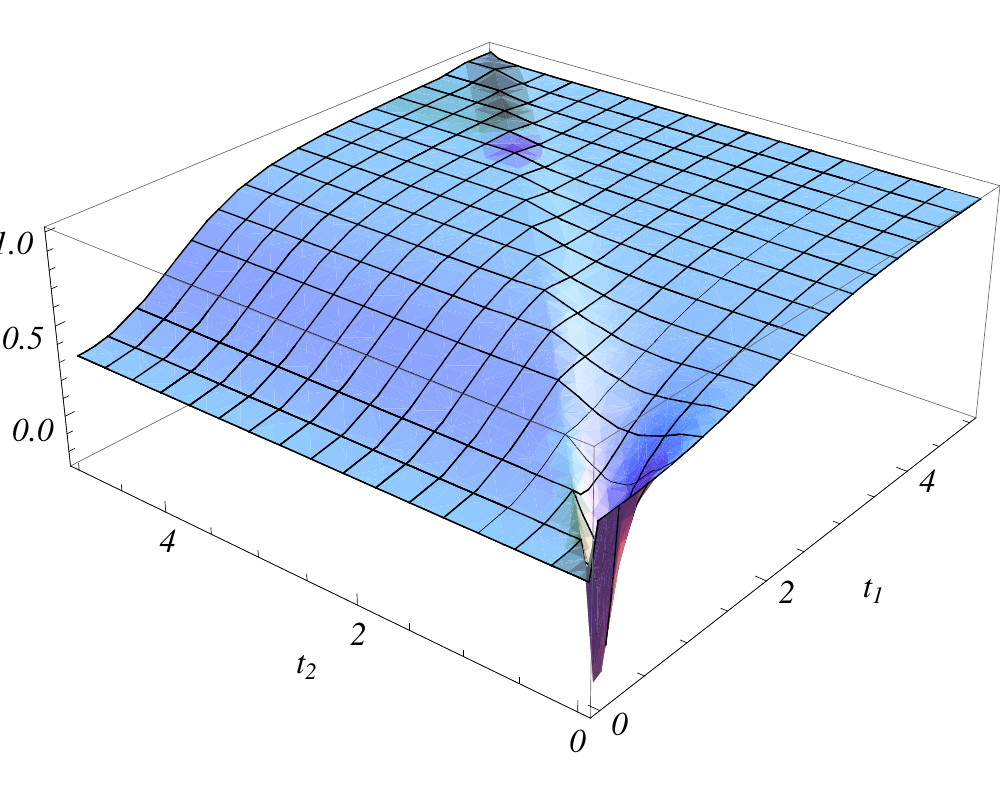}}
\caption{The GGT index of component $ 1 $ when $ t_1 \leq t_2 $ and $ t_2=0.1, 0.5, 1$ (left) and  $ t_1 > t_2 $ (right), for the distribution given in Example \ref{exmp4} (II).}
\end{center}
\end{figure} \label{fig13}

Figure 15 shows that in case (II) the GGT index of component $ 1 $ is positive for large values of $ t_1 $. 
Generally, while both components are alive the GGT index is smoothly increasing as $ t_1 $ is increased, 
though there will be a change in the GGT trend at the failure of one of the components. 
Under the assumption that $ t_2 < t_1 $, the GGT index is positive and increased as $ t_1 $ and $ t_2 $ 
gain larger values.

\begin{figure}[ht] 
\begin{center}
{\includegraphics[scale=0.55]{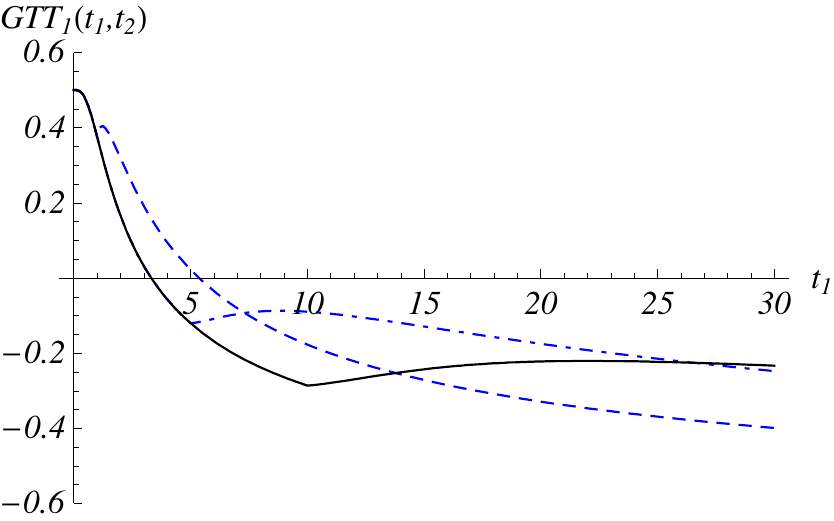}}
{\includegraphics[scale=0.55]{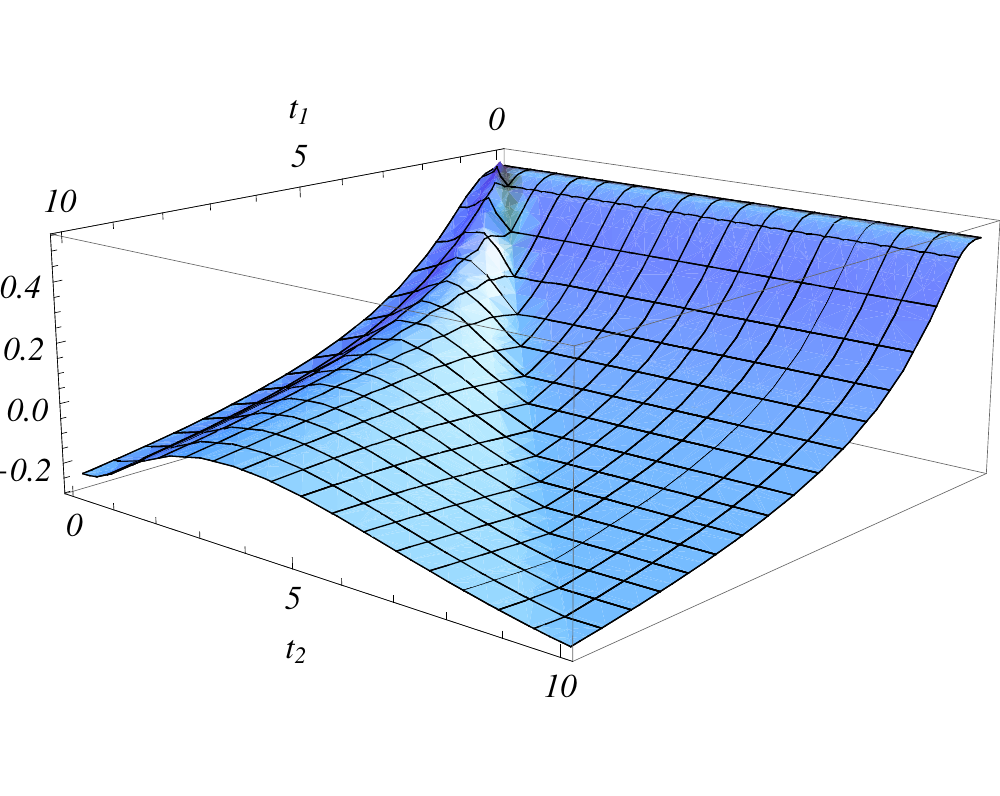}}
\caption{The GGT index of component $ 1 $ when $ t_1 \leq t_2 $ and $ t_2=1, 5, 10$ (left) and  $ t_1 > t_2 $ (right), for the distribution given in Example \ref{exmp4} (III).}
\end{center}
\end{figure} \label{fig14}

For the case (III), Figure 16 illustrates that the GGT index can receive positive and negative values in both assumptions. It is decreasing as long as both components are working. And when failure of component 2 is supposed to occur earlier, \textit{i.e.} $ t_2 < t_1 $, it is decreasing for large values of $ t_1 $ and $ t_2 $.
\par
For other properties of the aforementioned bivariate distributions related to bivariate IFR, see \cite{Bassan 2002}.
\end{example}

The following definition of stochastic order, which is in analogy with Definition \ref{df2}, 
grants a new tool for comparing random vectors based on the GGT indexes. 

\begin{definition}
Let $ (X_1,X_2) $ and $ (Y_1,Y_2) $ be two random vectors,  
each vector denoting the lifetimes of a pair of components working in the same environment. We say that
\begin{itemize}
\item  $ X_1 $ is less than $ Y_1 $ in terms of GGT index, and write $  X_{1} \leq_{GGT_1} Y_{1} $, if 
$$
GGT_{X_1}(t_1,t_2) \leq GGT_{Y_1}(t_1,t_2) \hbox{ for all $t_1, t_2 \in D^1$.}
$$

\item When $ t_1 > t_2 $,  we say that the residual  
lifetime of $ X_1 $ is less than the residual lifetime of $ Y_1 $ 
based on the conditioned GGT index, and write $  X_{1} \leq_{GGT_{1 | 2}} Y_{1} $, if 
$$
GGT_{X_{1} | X_{2}}(t_1 \, | \, t_2) \leq GGT_{Y_{1} | Y_{2}}(t_1 \ | \ t_2), \hbox{ for all $t_1, t_2 \in D^1_{t_2}$},
$$
where $ D^1_{t_2}=D^1_{X_1 | t_2} \cap D^1_{Y_1 | t_2} $.
\end{itemize}
 \end{definition}
 
It is not hard to prove the reflexivity and transitivity properties for the aforementioned orders. However, antisymmetry strictly depends on the cumulative hazard and might not be fulfilled in some cases. 
 
More detailed results on this notions will be considered elsewhere.

\subsection{Multivariate Case}
Assume that $ \textbf{T}=(T_1, \dots, T_n) $ is a non-negative  random vector, 
with a.c.\ joint distribution function and continuous joint d.f., 
representing the lifetime of $ n $ components. Let $ \textbf{e}=(1,\dots,1) $ and 
$ \textbf{t}=(t_1, \dots, t_n) $. Moreover, let $ I \subseteq \{1,\dots,n\} $ be the set of live components and, therefore, $ \bar{I} $ is the set of already failed components. Suppose that at time $ t $ it is known that $ \textbf{T}_{\bar{I}}=\textbf{t}_{\bar{I}} \leq t \textbf{e} $ and that $ \textbf{T}_I > t \textbf{e}$. The multivariate conditional hazard rate is defined as follows. 
\begin{definition} \label{def9bis}
For $ t>0 $, $ I \subseteq \{1, \dots, n\} $, and $ \textbf{t}_{\bar{I}} \leq t \textbf{e} $, the multivariate conditional hazard rate function of component $ i \in I$ is
\begin{equation} \label{eq23}
h_{i | \bar{I}}(t \,|\, \textbf{T}_{\bar{I}}=\textbf{t}_{\bar{I}}, \textbf{T}_I > t \textbf{e})
=\lim_{ \Delta t \downarrow 0} \dfrac{\mathbb P(t < T_i \leq t+ \Delta t ~|~\textbf{T}_{\bar{I}}=\textbf{t}_{\bar{I}},\textbf{T}_I > t \textbf{e})}{ \Delta t}.
\end{equation}
When $ \bar{I}=\varnothing $, the function in \eqref{eq23} reduces into 
\begin{equation} \label{eq24}
h_{i}(t \,|\, \textbf{T} > t \textbf{e})=\lim_{ \Delta t \downarrow 0} 
\dfrac{\mathbb P(t < T_i \leq t +  \Delta t ~|~ \textbf{T}>t \textbf{e})}{ \Delta t}.
\end{equation}
\end{definition}
The accumulated hazards can be defined as well.
\begin{definition}\label{def10}
For $ I \subseteq \{1,\dots,n\} $, $ i \in I $, $ \textbf{t}_{\bar{I}} > 0 \textbf{e} $, and $ t > 0 $, the accumulated hazard for the component $ i $ over interval $ (\max_{j\in \bar{I}} t_j,\max_{j\in\bar{I}} t_j +t) $ is 
acquired as below, 
\begin{equation}\label{eq25}
H_{i | \bar{I}} ( t ~|~ \textbf{t}_{\bar{I}} ) := \int^{\max_{j\in\bar{I}} t_j +t}_{\max_{j\in \bar{I}} t_j} h_{i|\bar{I}} ( u~|~\textbf{t}_{\bar{I}} ) du.
\end{equation}
When $ \bar{I}=\varnothing $ the quantity $ H_{i | \bar{I}} ( t~|~ \textbf{t}_{\bar{I}} ) $ will be simply denoted by $ H_i(t) $.
\end{definition}

Let $ t>0 $ be  fixed and assume that $ T_{j_1},\dots, T_{j_{k-1}}$, $ k>1 $, are the already failed components lifetimes at times $  t_{j_1}\leq \dots \leq t_{j_{k-1}} $, respectively, and also suppose that the rest of components are alive at time $ t $. For $ i \notin \{j_1,\dots, j_{k-1}\}$, the total hazard accumulated by component $ i $ is denoted by, 
\begin{equation}\label{eq26}
   \begin{aligned}
    &\hspace{-1cm}\Lambda_{i |  j_1,\dots , j_{k-1}} (t  \,| \, t_{j_1},\dots, t_{j_{k-1}}) \\
    &:= H_i (t_{j_1})+\sum^{k-1}_{l=2} H_{i | j_1,\dots , j_{l-1}}(t_{j_1}-t_{j_{l-1}}\,|\,t_{j_1}, \dots , t_{j_{l-1}}) \\
&+ H_{i | j_1,\dots , j_{k-1}} (t-t_{j_{k-1}}\,|\,t_{j_1}, \dots ,t_{j_{k-1}}).
  \end{aligned}
\end{equation}
For the case of $ k=1 $ we denote $\Lambda _i(t)\equiv H_i(t)$, $t \geq 0$. 
For more discussions about the properties and applications of the multivariate conditional 
hazard rate function, 
see  \cite{Shaked and Shanthikumar 1987} and \cite{Shaked and Shanthikumar 2014}. 
\par
The multivariate GGT index is defined by extension of the Definition \ref{def8} and \eqref{eq26} and is expressed as follows, where
\[
 D^1_{ X_i \ | \ (X_{j_1},\dots, X_{j_{k-1}}) }=
 \{  t >( t_{j_1},\dots, t_{j_{k-1}}) :  0< -\frac{\partial^k}{\partial t_{j_1},\dots, \partial t_{j_{k-1}}} \bar{F}(t, t_{j_1},\dots, t_{j_{k-1}})<1 \}. 
\]
\begin{definition}\label{def11}
If $ t_{j_1},\dots, t_{j_{k-1}}$, $ k>1 $, are the already failed components lifetimes at times $  t_{j_1}\leq \dots \leq t_{j_{k-1}}$, then for $ i \notin \{j_1,\dots, j_{k-1}\}$ the GGT index of component $i$ in $ (0,t_i) $ is given by
\begin{equation}\label{eq12}
\begin{aligned}
GGT_i(\textbf{t})&=
1-\frac{2 \int^{t_i}_0 \Lambda_{i | j_1,\dots , j_{k-1}} (u \,|\, t_{j_1},\dots, t_{j_{k-1}}) du}{t_i \Lambda_{i | j_1,\dots , j_{k-1}} (t_i \,|\, t_{j_1},\dots, t_{j_{k-1}})}, \qquad t_i \in D^1_{ X_i \ | \ (X_{j_1},\dots, X_{j_{k-1}}) },
\end{aligned}
\end{equation}
which depends on actual value of $ t_i $.
\end{definition}

%
\section{Vector Gini-type index}\label{sec8}
The notions which have been exploited in the previous section involve conditioning in multivariate setting, and are finalized to investigate ageing properties of dependent random lifetimes. 
It has been shown that a generalized version of the GT index is useful in this respect. 
Nevertheless, this section is devoted to introduce a vector GT index, which involves the notion of multivariate vector hazard rate.  
\par
According to this aim, let us recall that
Johnson and Kotz \cite{Johnson and Kotz 1975} defined a vector of multivariate failure rate and concepts of increasing and decreasing multivariate failure rate distributions. Specifically, 
let $ \textbf{X}=(X_1, \dots, X_n) $ be a non-negative a.c.\ random vector denoting the lifetime of $ n $ components, 
and assume the joint survival function as  
$\bar{F}_{\textbf{X}}(\textbf{t})=P(X_1 > t_1, \dots, X_n > t_n)$.  
The corresponding multivariate vector hazard rate is given as
\begin{equation}
\begin{aligned}
h_{\textbf{X}}(\textbf{t})&=-\nabla \log (\bar{F}_{\textbf{X}}(\textbf{t})) \\
&=(-\frac{\partial}{\partial t_1}, \dots, -\frac{\partial}{\partial t_n})\log (\bar{F}_{\textbf{X}}(\textbf{t})) \\
&=( h_{\textbf{X}}(\textbf{t})_1, \dots,  h_{\textbf{X}}(\textbf{t})_n),
\end{aligned}
\end{equation}
in which
\[
   h_{\textbf{X}}(\textbf{t})_i=-\frac{\partial}{\partial t_i} \log (\bar{F}_{\textbf{X}}(\textbf{t})), \qquad i=1, 2, \dots, n,
\]
is the hazard rate of component $ i $-th while other components are working. Thus, the cumulative multivariate hazard rate of $i $-th component up to time $ t_i$ is derived as

\begin{equation}
H_{\textbf{X}}(\textbf{t})_i=\int^{t_i}_0 h_{\textbf{X}}(t_1,...,t_{i-1}, u, t_{i+1}, ...,t_n )_i du.
\end{equation}
Also, the multivariate vector of cumulative hazard rate is 
\[
H_{\textbf{X}}(\textbf{t})=(H_{\textbf{X}}(\textbf{t})_1, \dots, H_{\textbf{X}}(\textbf{t})_n).
\]
The following definition presents the notion of vector Gini-type (VGT) index of component $ i $-th, $ i=1, 2, \dots, n $.

\begin{definition}
The VGT of non-negative random variable $ X_i $, $ i=1, 2, \dots, n $ in time interval $ (0,t_i] $ is

\begin{equation}
VGT_{\textbf{X}}(\textbf{t})_i = 1-\dfrac{2 \int^{t_i}_{0} H_{\textbf{X}}(t_1,...,t_{i-1}, u, t_{i+1}, ...,t_n )_i du}{t_i H_{\textbf{X}}(\textbf{t})_i}, \qquad t \in D^1_i,
\end{equation}

where $ D^1_i=\bigcap^n_{i=1}D^1_{ X_i \ | \ (X_{j_1},\dots, X_{j_{k-1}}) } $.
\par 

Therefore, the vector of MGT index is 
\[
VGT_{\textbf{X}}(\textbf{t})=(MGT_{\textbf{X}}(\textbf{t})_1, \dots, VGT_{\textbf{X}}(\textbf{t})_n), \qquad \forall t_i \in D^1_i,\ i=1, \dots, n.
\]
\end{definition}

\begin{remark}
If $ X_1, \dots, X_n $ are mutually independent then 
\[
 h_{\textbf{X}}(\textbf{t})_j= h_{X_i}(t_i).
\] 
Consequently, 
\[
VGT_{\textbf{X}}(\textbf{t})_i=GT_{X_i}(t_i) .
 \]
\end{remark}

We recall that, according to \cite{Johnson and Kotz 1975}, if $  h_{\textbf{X}}(\textbf{t})_i $ is an increasing (decreasing) function of $ t_i$, for all $ i=1, 2, \dots, n $, then the distribution is called vector multivariate IFR (DFR).

\par
Finally, we point out that the generalization of GT index has been discussed in this paper as a tool for comparing a component lifetime property while it depends on other components. This is due to the fact that failure of one or more components, exposes the failure rate of the remained components to vary. 
Besides, the vector of all GGT indexes can lead one to the properties of all components in a system.  
\section{Conclusions}
So far in the reliability literature, different ageing concepts have been defined to declare the properties of systems. The GT index is remarkably applicable to diagnose the ageing properties of complex systems. 
\par
According to system models that we have studied in this paper, the ageing of series system is faster than its own components. Considering the GT index of more complicated systems, a parallel-series system with shared components deteriorates faster than its dual series-parallel system and this property is enhanced by increasing the number of components. 
\par
The GGT index has been utilized for comparing the properties of a group of dependent lifetime variables to another competitor group. Besides, by the conditioned GGT index we have been capable of comparing ageing properties of functioning units while assuming the earlier failure of some other units. This is due to the fact that failure of one or more units, exposes the failure rate of the remained components to vary. 

\section*{Acknowledgements}
We highly appreciate the support of Iranian Ministry of Science and Technology, 
and of the Italian group GNCS of INdAM. Authors are also grateful for the collaboration 
of Ordered and Spatial Data Center of Excellence at Ferdowsi University of Mashhad, Iran, 
with Department of Mathematics at University of Salerno, Italy. 
\par
The constructive criticism of two anonymous reviewers is gratefully acknowledged. 

\end{document}